\documentclass[12pt,preprint]{aastex}
\usepackage{graphicx}

\shorttitle{Optical Monitoring of S5~0716+714}
\shortauthors{Wu et al.}

\received{}
\accepted{}

\begin{document}

\title{Simultaneous $B'V'R'$ Monitoring of BL Lacertae Object S5~0716+714
       and Detection of Inter-Band Time Delay}

\author{Jianghua Wu}
 \affil{Department of Astronomy, Beijing Normal University, Beijing 100875,
        China}
 \email{jhwu@bnu.edu.cn}

\author{Markus B\"ottcher}
 \affil{Astrophysical Institute, Department of Physics and Astronomy,
        Ohio University}
 \affil{Athens, OH 45701, USA}

\author{Xu Zhou}
 \affil{Key Laboratory of Optical Astronomy, National Astronomical
        Observatories, Chinese Academy of Sciences}
 \affil{20A Datun Road, Chaoyang District, Beijing 100012, China}

\author{Xiangtao He}
 \affil{Department of Astronomy, Beijing Normal University, Beijing 100875,
        China}

\author{Jun Ma, and Zhaoji Jiang}
 \affil{Key Laboratory of Optical Astronomy, National Astronomical
        Observatories, Chinese Academy of Sciences}
 \affil{20A Datun Road, Chaoyang District, Beijing 100012, China}

\begin{abstract}
We present the results of our optical monitoring of the BL Lac object
S5~0716+714 on seven nights in 2006 December. The monitoring was carried out
simultaneously at three optical wavelengths with a novel photometric system.
The object did not show large-amplitude internight variations during this
period. Intranight variations were observed on four nights and probably on
one more. Strong bluer-when-brighter chromatism was detected on both
intranight and internight timescales. The intranight variation amplitude
decreases in the wavelength sequence of $B'$, $R'$, and $V'$. Cross
correlation analyses revealed that the variability at the $B'$ and $V'$ bands
lead that at the $R'$ band by about 30 minutes on one night.
\end{abstract}

\keywords{BL Lacertae objects: individual (S5~0716+714) --- galaxies: active
--- galaxies: photometry}

\section{INTRODUCTION}
Blazars represent the most violently variable objects among all active galactic 
nuclei (AGNs). They show rapid and strong variability, high and variable 
polarization, and a non-thermal continuum. The variable continuum is believed 
to come from near the base of the relativistic jet closely aligned with
our light of sight. The jet is probably powered and accelerated by a rotating
supermassive black hole surrounded by an accretion disk. A blazar can either
be termed as a flat-spectrum radio quasar (FSRQ) or a BL Lac object, depending
on whether or not it shows strong emission lines in its spectrum.

The spectral energy distribution (SED) of blazars has a two-bumped structure:
The low-frequency bump ranges from radio to UV or X-ray frequencies and is
believed to come from the synchrotron emission from the relativistic electrons
in the jet. The high-frequency bump, from X-rays to $\gamma$-rays, is usually
interpreted as emission from inverse Compton scattering of the low frequency
radiation by the same ensemble of relativistic electrons responsible for
the synchrotron emission \citep[for a recent review, see][]{bottcher07a}.
According to the peak frequency of their synchrotron emission, BL Lac objects
can be further divided into low-frequency-peaked BL Lacs (LBL) and
high-frequency-peaked BL Lacs (HBL). The former subclass has the synchrotron
peak at IR--optical wavelengths while the latter subclass has the peak in the
UV or soft X-ray band.

The BL Lac object S5 0716+714, probably at a redshift of 0.31$\pm$0.08
\citep{nilsson08}, is one of the best studied blazars. It is highly variable
from radio to X-ray \citep{wagner96} and $\gamma$-ray wavelengths
\citep{lin95,chen08}. Most recently, MAGIC detected very high energy
$\gamma$-ray emission from this object\citep{anderhub09}. Its SED has been
observed and studied by a number of multi-wavelength campaigns \citep[e.g.,]
[]{giommi99,giommi08, taglia03,ferrero06,foschini06,ostorero06,villata08}.
The majority of the observations revealed a two-bumped structure of the SED
with a concave shape at 2 -- 3 keV. Furthermore, it was found that its
emission from radio to soft X-rays, likely due to the synchrotron process,
could be highly variable, while the hard X-ray (from 3 to 10 keV) flux, most
probably dominated by inverse Compton emission, usually remained
constant.\footnote{An exception was found by \citet{ferrero06}, who detected
its delayed hard X-ray variation as well.} A model with two synchrotron
self-Compton (SSC) components was proposed to explain its SED changes, with
one component highly variable and the other constant \citep{chen08,giommi08}.

In the optical regime, S5~0716+714 is one of the brightest blazars. It has 
a duty cycle close to 1, which means that it almost never stops its variation
\citep[e.g.,][]{wagner95,stalin09,chandra11}. Its optical variability has been
studied on different timescales by a number of authors \citep{ghise97,sagar99,
villata00,nesci02,nesci05,qian02,raiteri03,xie04,wu05,wu07a,montagni06,
stalin06,pollock07,gupta08,sasada08,zhang08,stalin09,poon09,zhang10a,carini11,
chandra11}. Its variation rate can be faster than 0.10-0.12 $\rm{mag\,hr^{-1}}$
\citep{sagar99,villata00,wu05,montagni06}. A maximum rate of 0.16
$\rm{mag\,hr^{-1}}$ was reported by \citet{nesci02}. The variation timescale
can be as short as 15 minutes \citep{sasada08,rani10a,chandra11}. Most
recently, \citet{fan11} even reported a 0.611 mag variation over 3.6 minutes
in this object.

The color or spectral behavior of blazars is a subject of much debates.
Some authors found a bluer-when-brighter (BWB) chromatism \citep[e.g.,][]
{vagnetti03}, some others claimed the opposite behavior, namely, a
redder-when-brighter (RWB) trend \citep[e.g.,][]{ramirez04}, or no clear
tendency \citep[e.g.,][]{bottcher07b,bottcher09}. It appears that BL Lacs are
BWB while FSRQs are RWB \citep{fan00,gu06,hu06,rani10b}.\footnote{However,
\citet{gu11} found that only one FSRQ is RWB in a sample of 29 SDSS FSRQs.}
A certain object may display different trends in different variation modes
\citep[e.g.,][]{wu05,poon09} or on different timescales \citep[e.g.,][]
{ghise97,raiteri03}. One reason for these divergences is that almost all
previous observations were done {\sl quasi}-simultaneously at multiple
wavelengths. When the exposure was switched from one wavelength to the other,
the brightness of the source might have changed. So the quasi-simultaneous
observations cannot obtain real color, and the color behavior drawn from
these observations should be taken with caution.

We monitored S5~0716+714 for seven nights in 2006 December with a novel
optical photometric system \citep{wu07a,wu07b}. This system enables
simultaneous photometry at multiple optical wavelengths, and thus has the
advantages to accurately trace the color change of blazars and to increase
the temporal resolution at individual wavelengths. The
resulting data were used to study the properties of the short-term
variability of S5~0716+714, to analyze its color behavior during variations,
and to search for the correlation and time lags between the variations at
various optical wavelengths. Section 2 describes our observations and data
reduction procedures. Section 3 presents the light curves and the results
of the color and correlation analyses. The conclusions and discussions are
given in Section 4.

\section{OBSERVATIONS AND DATA REDUCTION}
The monitoring of S5~0716+714 was carried out at Xinglong Station of the
National Astronomical Observatories of China in 2006 December with a novel
photometric system, which consists of a 60/90 cm Schmidt telescope, an
objective prism, a CCD camera, and a multi-peak interference filter (MPIF).
Beams of light in multiple passbands are refracted differentially by the
objective prism, reflected by the spherical main mirror, pass through the
MPIF simultaneously, and are focused on the CCD separately. So we have
multiple ``images" on the frame for each object. The left three passbands
of the MPIF are close to the broadband $B$, $V$, and $R$, so we designate
them as $B'$, $V'$, and $R'$. A fourth passband beyond 9400 {\AA} is
neglected, because the CCD has a very low response at those wavelengths.
For details of the system, see \citet{wu07a,wu07b}. 

A change to the system is that we now use a new $4096\times4096$ E2V CCD.
Compared to the old one, it has a smaller pixel size of $12\,\micron$ and a
higher spatial resolution of $1.3\,\arcsec\rm{pixel}^{-1}$. When observing
blazars, we read out only the central $512\times512$ pixels as a frame,
which has a field of view of about $11'\times11'$. The key advantage of the 
new CCD is its high quantum efficiency at blue wavelengths (92.2\% at 
4000 {\AA}). So we now have useful data in one more band, the $B'$ band, 
in addition to the $V'$ and $R'$ bands for the old system \citep{wu07a,wu07b}.

The monitoring covered the period from 2006 December 18 to 27 (from JD
2,454,088 to 2,454,097), with no observation on December 19, 25, and 26 due
to bad weather. The exposure times are all 150 seconds. Taking into account
the readout time of about 6 seconds, we achieved a temporal resolution of
less than three minutes in three passbands. The average FWHMs of the stellar
images are 3\farcs1, 3\farcs1, 3\farcs6, 4\farcs3, 3\farcs4, 3\farcs7, and
5\farcs4 on JDs~2,454,088, 2,454,090, 2,454,091, 2,454,092, 2,454,093,
2,454,094, and 2,454,097, respectively.

\begin{figure}
\plotone{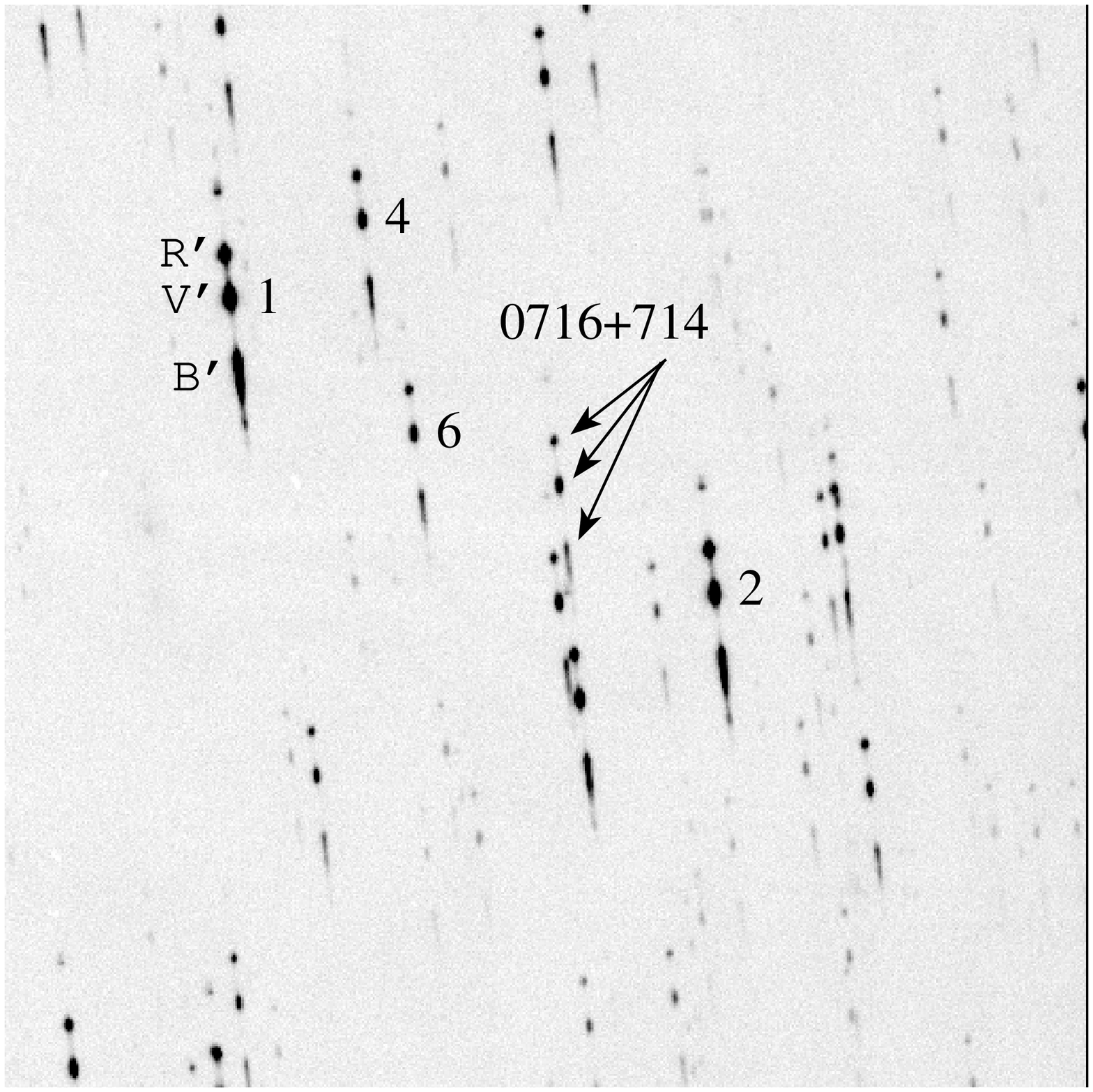}
\caption{An example frame taken with the novel photometric system. Labeled are
target blazar and four reference stars used for flux calibration. The $B'$,
$V'$, and $R'$ ``images" are marked for Star 1 for an example. North is to
the up and east, to the left.}
\label{F1}
\end{figure}

An example frame is shown in Figure~\ref{F1}. We rotated the objective prism
(and thus the dispersion direction) by a small angle from the standard vertical
position. As the result, some overlaps of images were resolved, especially the
potential overlap of the $B'$ image of S5~0716+714 with the $R'$ image of a
star below \citep[for a comparison, see Fig.~\ref{F1} in this paper and Fig.~2
of][]{wu07a}.

The data reduction included bias subtraction, flat-fielding, extraction of
instrumental aperture magnitude, and flux calibration. We defined a specially
shaped aperture, as in \citet{wu07a}, but with an update. In \cite{wu07a}, the
traditional circular aperture was equally split into two semicircles and a
rectangle was inserted between them. Then an elongated aperture was obtained
for the elongated images. For the inclined and elongated stellar images in
this paper, we replaced the rectangle with a parallelogram. The chords of the
two semicircles were attached to the top and bottom sides of the parallelogram,
respectively, while the left and right sides of the parallelogram were set
parallel to the dispersion line. Different parallelogram heights were adopted
for the apertures in different passbands. They are 11, 7, and 4 pixels for the
$B'$, $V'$, and $R'$ images, respectively. The corresponding parallelogram
widths (or the diameters of the semicircles) were set to be 5, 6, and 6
pixels, respectively. All sky annuli were set to have inner and outer radii
of 11 and 15 pixels, but an area within 6 pixels in right ascension to the
central axis of the aperture was excluded in order to eliminate the possible
contamination of the background by the apertures. Stars 1, 2, and 4 in
\citet{villata98} were used as reference stars. Their average brightness was
utilized to calibrate the flux of S5~0716+714. Star 6 is slightly brighter
than S5~0716+714 and was used as a check star \citep[for a reasonable
selection of reference and check stars, see][]{howell88}. The above aperture
sizes in the three bands were chosen based on several trials to minimize the
scatters in the differential magnitudes between the check star and the three
reference stars. As in \citet{wu07a}, the standard $B'$, $V'$, and $R'$
magnitudes of these four stars were mimicked by their broadband $B$, $V$, and
$R$ magnitudes. There should be some differences between these two magnitude
systems, but they are tiny \citep{wu07a}, and are irrelevant for the results
of our variability study presented here. The results 
are presented in Table~1. The columns are observation date and time (UT), 
Julian Date, exposure time, and magnitudes and errors of S5~0716+714 and 
differential magnitudes of star 6 in the $B'$, $V'$, and $R'$ bands, 
respectively.

\section{RESULTS}
\subsection{Light Curves}
\begin{figure}
\plotone{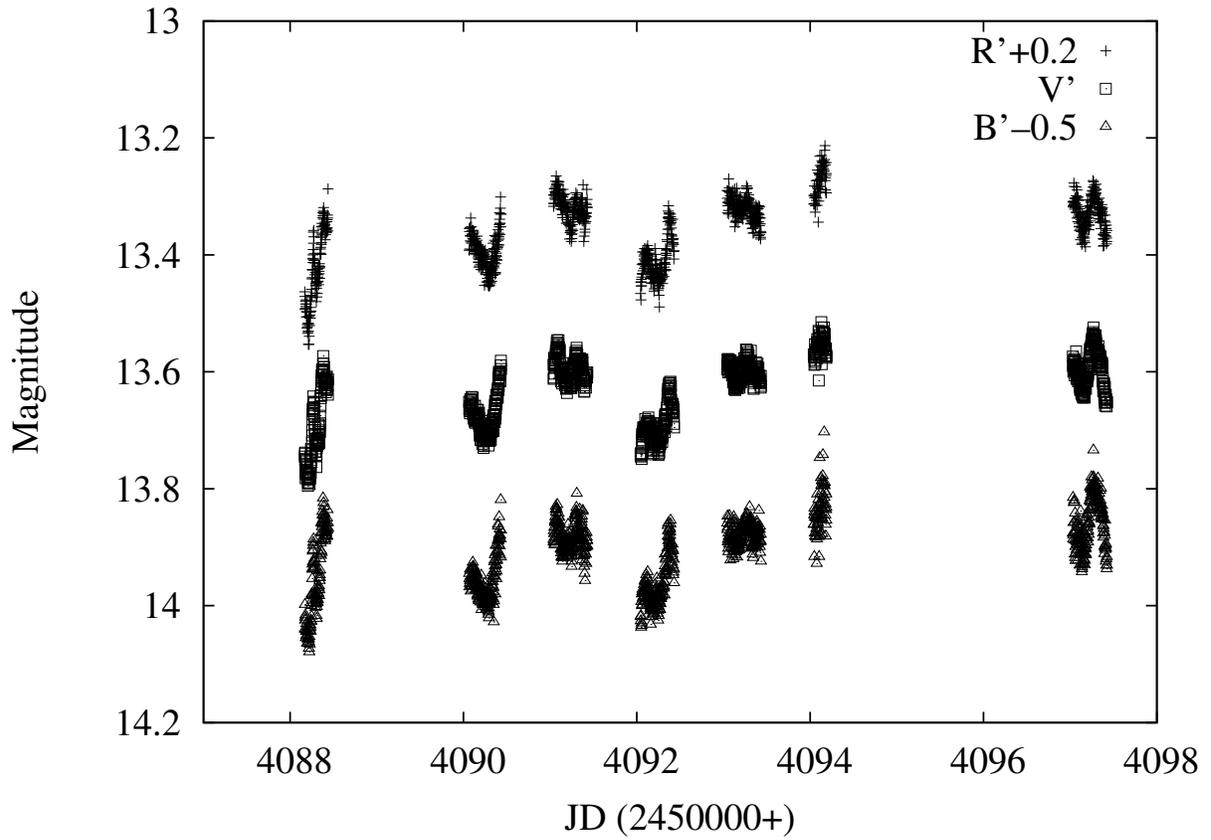}
\caption{Light curves in $B'$, $V'$, and $R'$ bands for the entire monitoring
period. For clarity, the $B'$ and $R'$ band light curves are shifted by 0.2
and $-0.5$ magnitudes, respectively.}
\label{F2}
\end{figure}

Figure~\ref{F2} shows the total light curves of S5~0716+714 in the $B'$, $V'$,
and $R'$ bands. The $B'$ and $R'$ band light curves are shifted by 0.2 and
$-0.5$ magnitudes, respectively, in order to make the variations clearly
visible. The nightly-averaged brightness of this object was quite stable in
this period and varies by less than 0.2 mags. On the other hand, intranight
variations can still be observed on most nights by visual inspection. 

We then performed a quantitative assessment on whether there are variations
on these seven nights. Here we used the one-way analysis of variance (ANOVA)
proposed by \citet{diego98} and \citet{diego10}. It is proved to be a
powerful and robust estimator for variations \citep{diego10}. In brief, the
observations on each night were divided into groups of five consecutive data
points. The variance of the means of each group about the overall mean was
computed, as was the mean of the variances or dispersions (residuals) within
each group. Then the ratio between these two variances was calculated and
multiplied by five, the number of observations in each group. The value
obtained behaves as the $F$ statistic. For a certain significance level, if
$F$ exceeds the critical value, the null hypothesis that there is no variation
will be rejected. The results are presented in Table~2. Column 1 is the
Julian Date. Columns 2 and 3 are the degrees of freedom for the groups and
residuals, respectively.  Columns 4, 5, and 6 are $F$ values in the $B'$,
$V'$, and $R'$ bands, respectively. Column 7 is the critical value at the
99\% significance level at the specified two degrees of freedom of Columns 2
and 3.  Column 8 indicates whether there is variation. As can be seen, there
are variations on JDs~2,454,088, 2,454,090, 2,454,092, and 2,454,097. The
$F_{\rm B'}$, $F_{\rm V'}$, and $F_{\rm R'}$ are all greater than the critical
values on the four nights. On one more night, JD~2,454,091, there may be
variation, too. The $F_{\rm V'}$ and $F_{\rm R'}$ are greater than the
critical value, whereas the $F_{\rm B'}$ is slightly smaller than it on that
night. On the two remaining nights, JDs~2,454,093 and 2,454,094, there is no
variation.

\begin{figure}
\epsscale{0.5}
\plotone{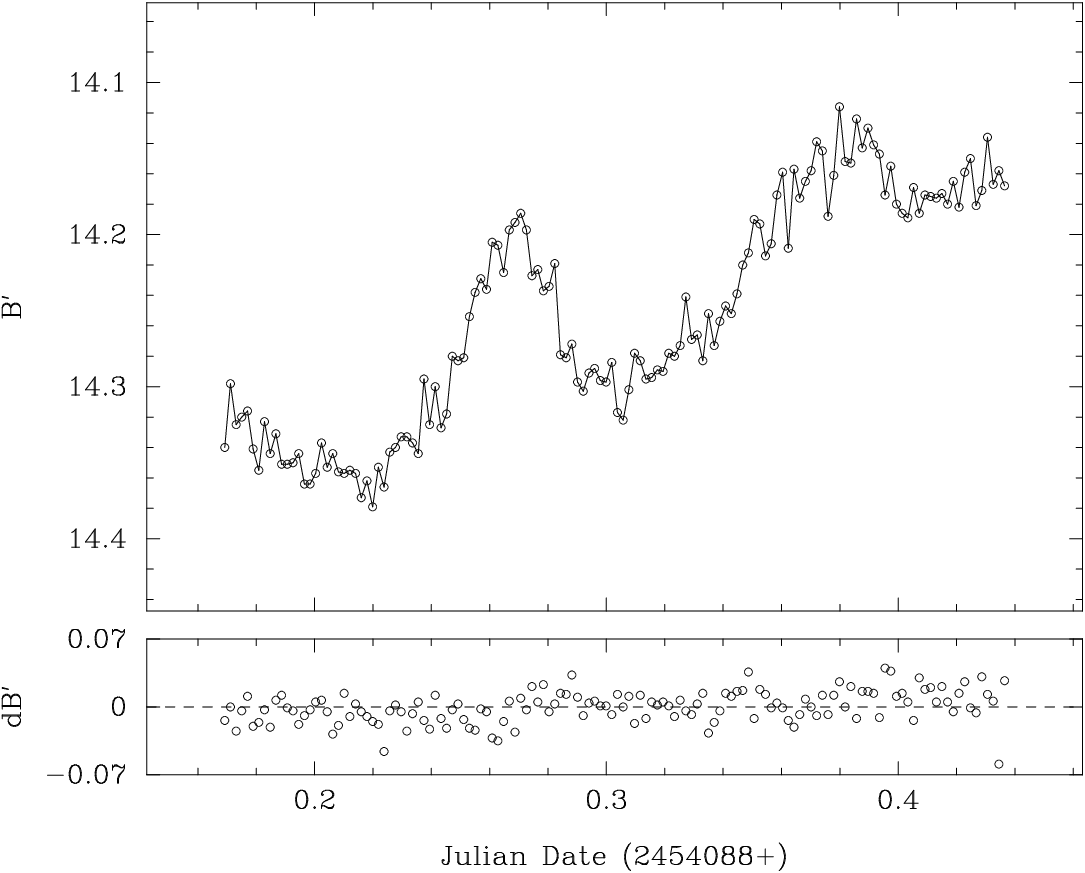}
\plotone{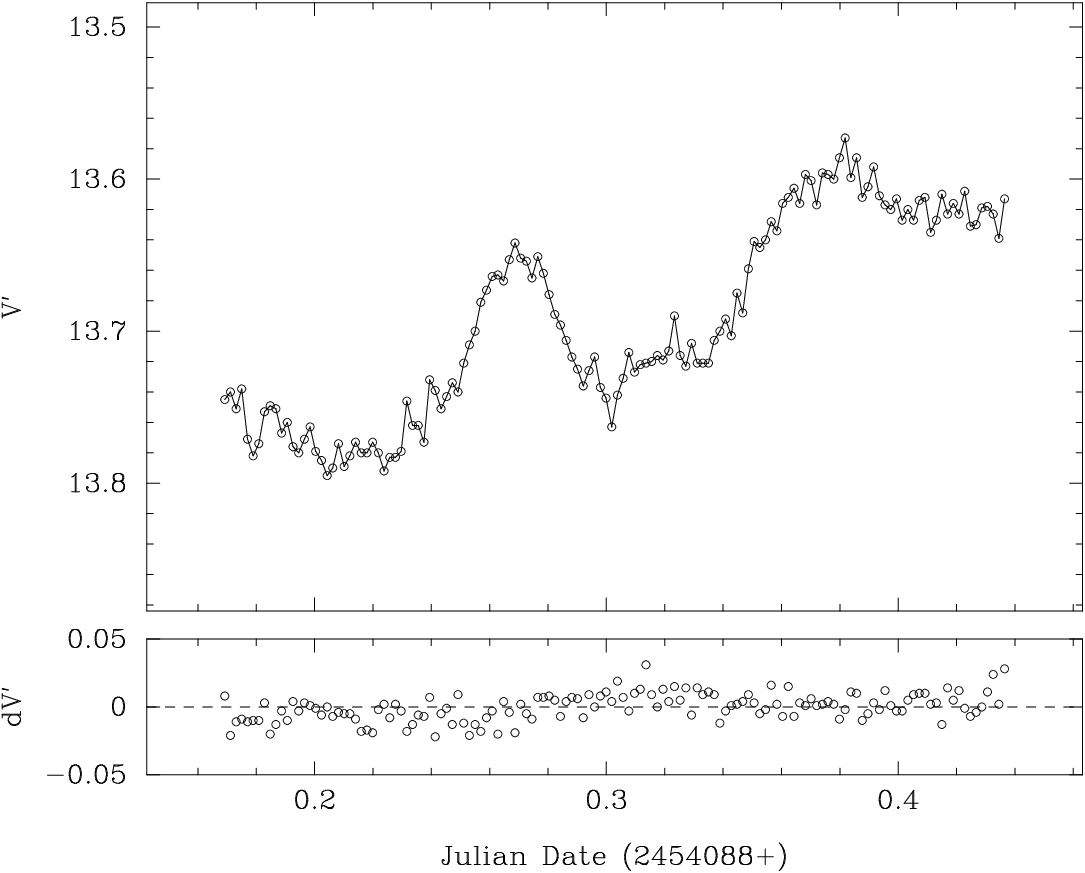}
\plotone{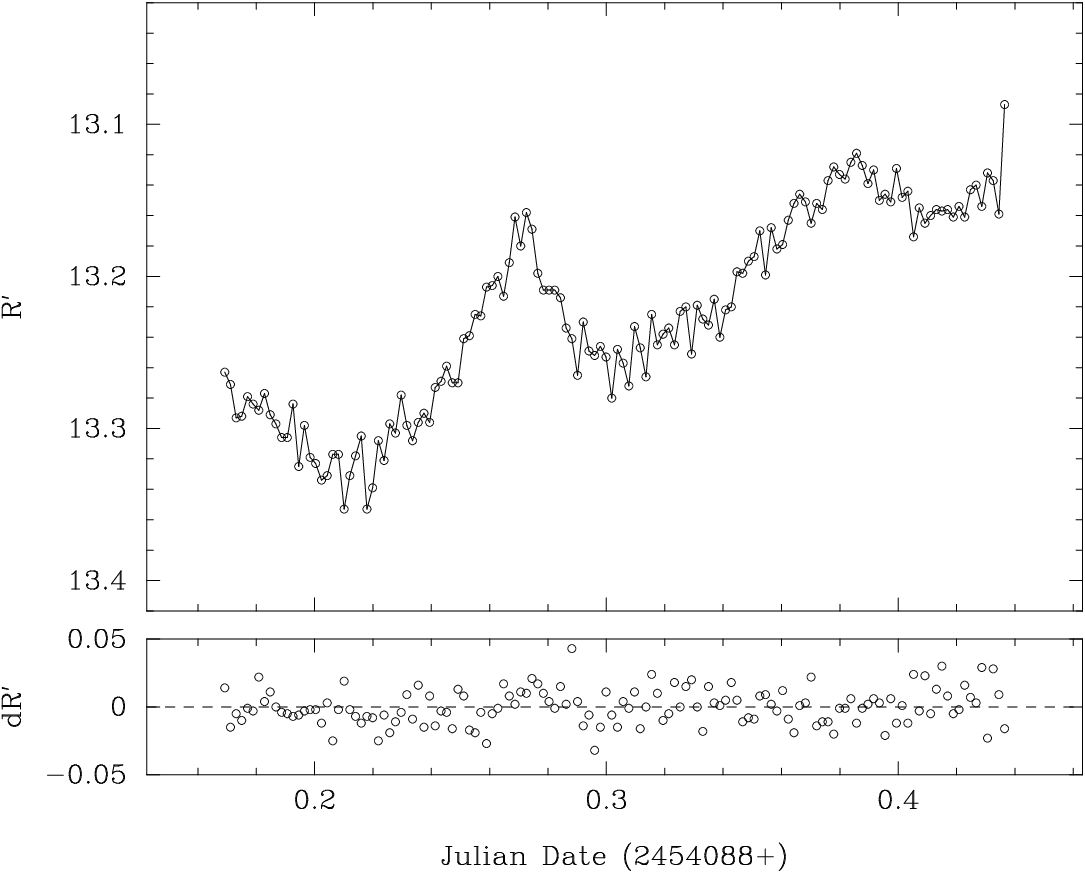}
\caption{Intranight light curves on JD~2,454,088 in the $B'$, $V'$, and $R'$
bands. The associated small panels give the differential light curves of the
check star (star 6).}
\label{F3}
\end{figure}

\begin{figure}
\plotone{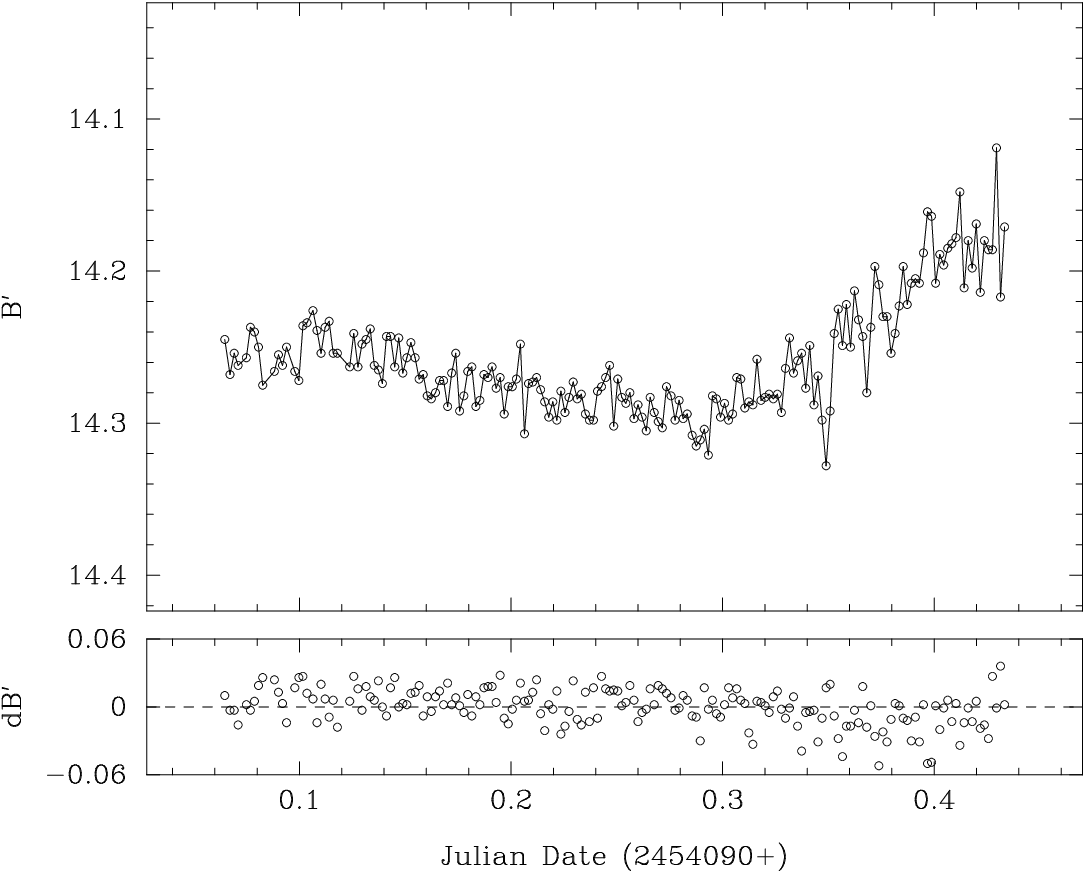}
\plotone{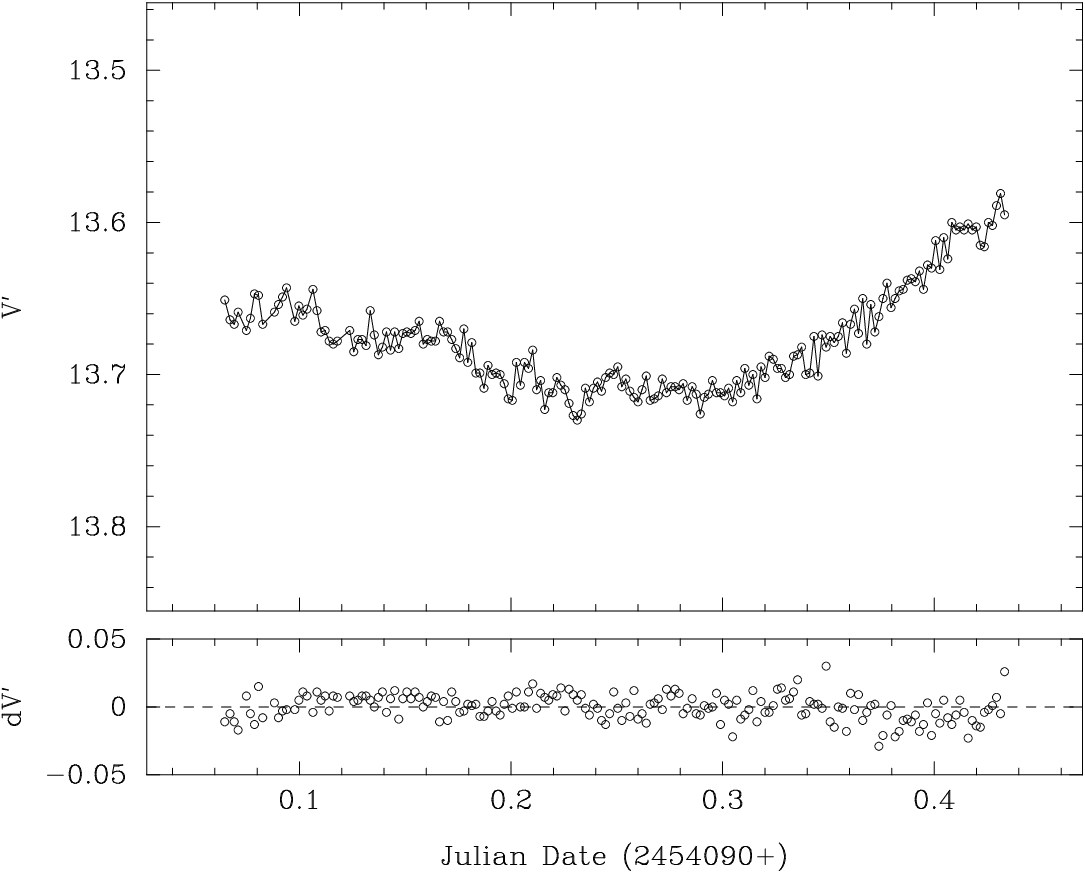}
\plotone{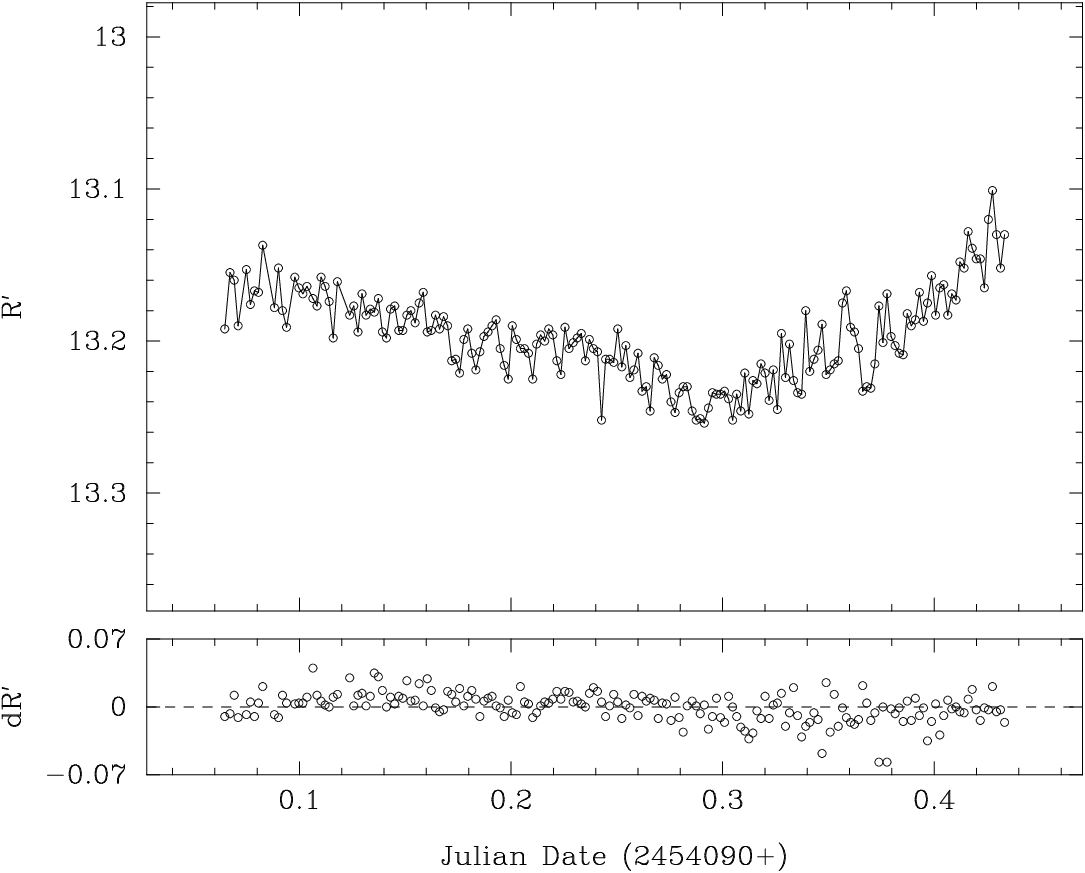}
\caption{Intranight light curves on JD~2,454,090 in the $B'$, $V'$, and $R'$
bands. The associated small panels give the differential light curves of the
check star (star 6).}
\label{F4}
\end{figure}

\begin{figure}
\plotone{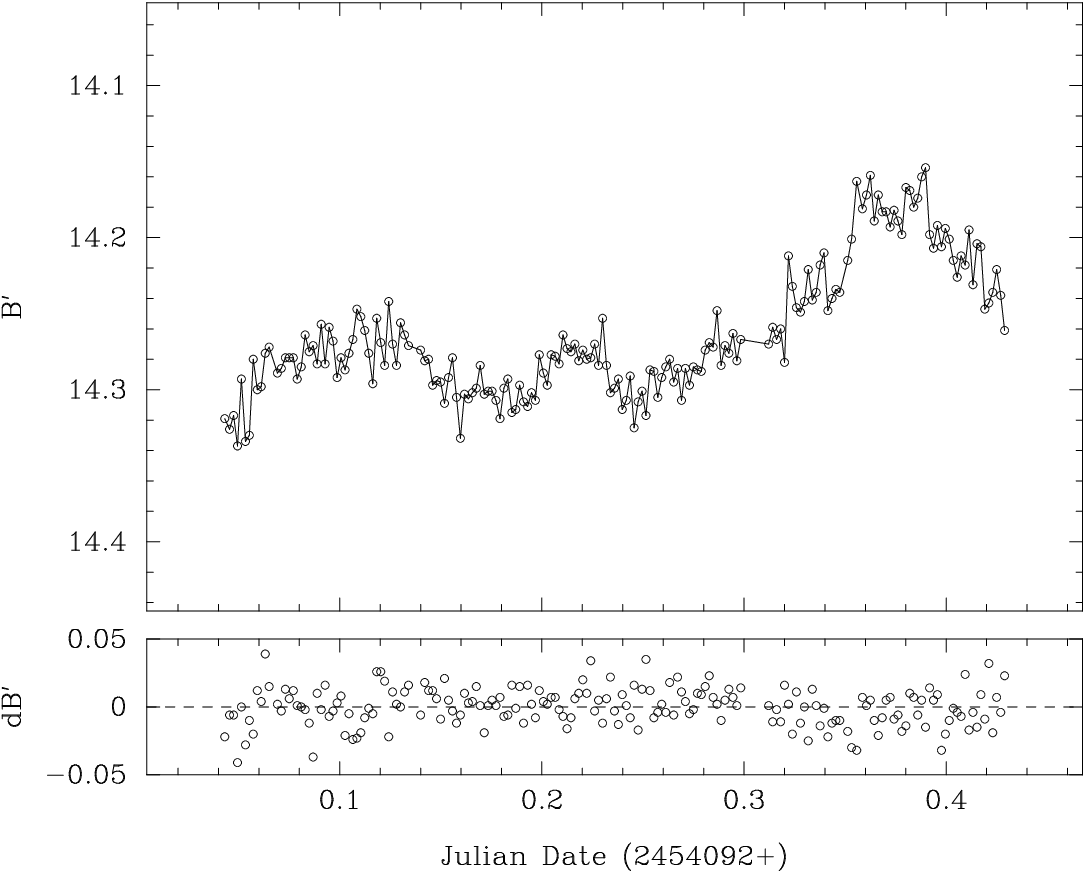}
\plotone{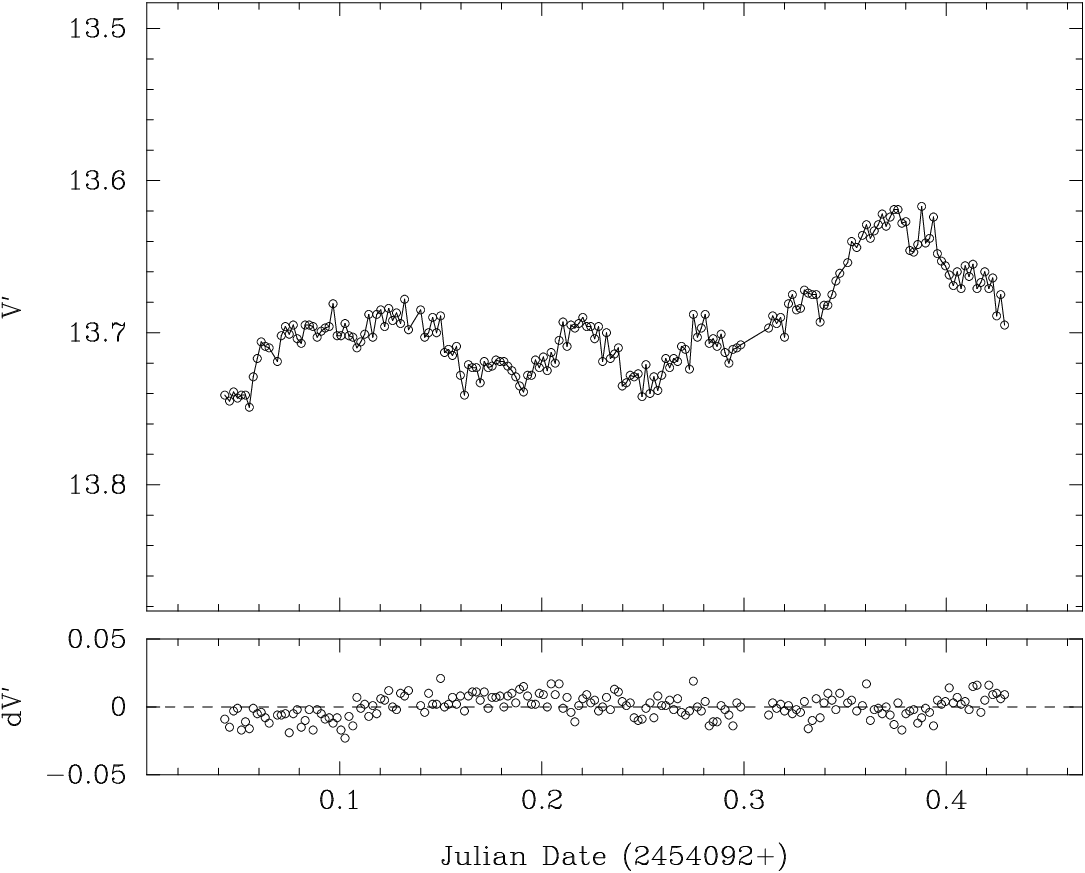}
\plotone{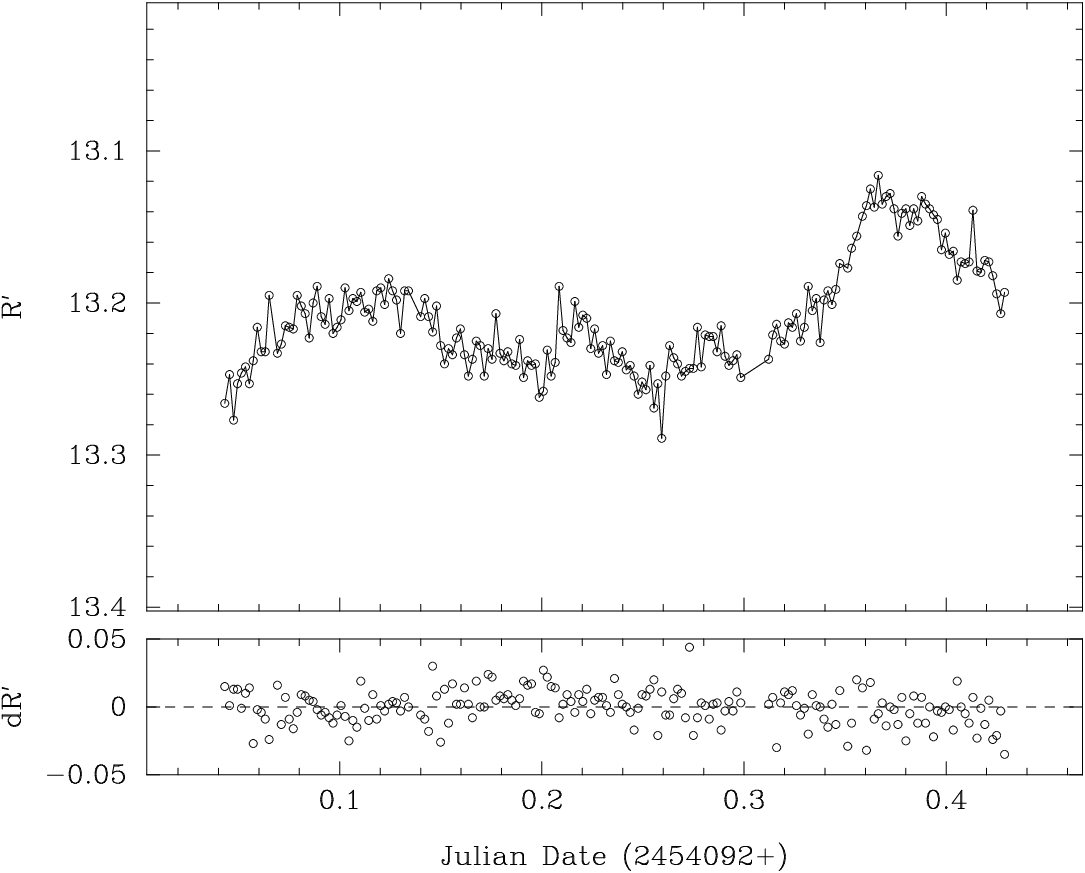}
\caption{Intranight light curves on JD~2,454,092 in the $B'$, $V'$, and $R'$
bands. The associated small panels give the differential light curves of the
check star (star 6).}
\label{F5}
\end{figure}

Figures~\ref{F3}, \ref{F4}, \ref{F5}, and \ref{F6} show the intranight light
curves of S5~0716+714 on the four nights with variations, JDs~2,454,088,
2,454,090, 2,454,092, and 2,454,097, respectively. The associated small panels
give the differential light curves of the check star (the means were set to 0).
For these four nights, we calculated the intranight variation amplitudes with
the definition of \citet{heidt96},
$A=\sqrt{(m_{\rm max}-m_{\rm min})^2-2\sigma^2}$, where $m_{\rm max}$ and
${m_{\rm min}}$ are the faintest and brightest magnitudes during any given
night, respectively and $\sigma$ is the standard deviation of the 'variation'
of the check star. The last point in the $R'$-band light curve on JD~2,454,088
and the point at (0.26764, 14.034) in the $B'$-band light curve on JD~2,454,097
are likely to be spurious measurements (see Figs.~\ref{F3} and \ref{F6}) and
were discarded before the calculations. The results are listed in Table~3.
It can be seen that the $B'$-band variations have the largest amplitudes and
the $V'$- and $R'$-band amplitudes are similar, with the latter lightly larger
than the former on three of the four nights. In the shock-in-jet model, the
variation amplitude decreases towards longer wavelengths. This has been
observed in S5~0716+714 \citep[e.g.,][]{wu07a,poon09,fan09} and in other
blazars \citep[e.g.,][]{nesci98,fan00,dai11}. However, the amplitudes may also
be related to the relative position of the frequency at which the observations
were made and the peak frequency of the synchrotron component in the SED of
blazars. The latter frequency itself may also shift from time to time
\citep[e.g.,][]{giommi99,giommi00,tavecchio01,zhang10b}. Taking these facts
into account, our results that the $R'$-band amplitudes are slightly larger
than the $V'$-band ones may not be unreasonable. As mentioned in \citet{wu07a}
and \citet{dai11}, the larger variation amplitude at higher frequency can
partly explain the BWB chromatism observed in blazars.

\subsection{Color-Magnitude Correlation}
Controversy exists in the studies of the spectral or color behavior of
blazars, as mentioned in \S1. For our data on the four nights with
variations, JDs~2,454,088, 2,454,090, 2,454,092, and
2,454,097, the color indices $B'-V'$ and $B'-R'$ were calculated and plotted
against the $B'$ magnitudes in the left and right panels of Figure~\ref{F7},
respectively. Different symbols mark data on different nights. They were
fitted linearly and separately for each single night. We used the BCES
regression \citep[bivariate correlated errors and intrinsic scatters,][]
{akritas96} to do the fitting. The BCES regression has the advantage of taking
into consideration not only the measurement errors, but also the intrinsic
scatters. The regression lines are plotted in Figure~\ref{F7} with different
styles for different nights. The thick solid lines are the fittings to all
points. The correlation coefficients are given at the upper-left corners.
It can be seen that the object shows strong BWB chromatisms on both intranight
and internight timescales. This result is in agreement with several previous
results \citep{wu05,wu07a,hao10,chandra11} and with the short-term behavior
obtained by \citet{ghise97} and \citet{zhang08}. The much larger correlation
coefficient in the right panel indicates a stronger correlation between
$B'-R'$ and $B'$ than between $B'-V'$ and $B'$. A similar result has been
reported by \citet{raiteri03}, who found a stronger BWB trends with the
$B-I$ color than with the $B-R$ color.

In addition to the flatter-when-brighter trend during rapid flares,
\citet{ghise97} have reported that the correlation was rather insensitive
for the long-term variations of S5~0716+714. A similar phenomenon was also
recognized by \citet{raiteri03} in the same object. Our monitoring lasted
for only ten days, so we cannot address the long-term color behavior of
this object.

\subsection{Color Evolution During Flares}
In a variety of blazar emission models, one expects an interplay between
gradual particle acceleration, radiative cooling and escape. This will lead
to a loop-like path of the blazar's state in a color-magnitude (or spectral
index-flux) diagram. The
direction of this spectral hysteresis can be either clockwise or anticlockwise,
depending on the relative values of the acceleration, cooling and escape
timescales \citep{cg99,dermer98}, as well as the frequency at which the
observation is made relative to the peak frequency of the synchrotron
component in the SED of the blazar \citep[see Figs. 3 and 4 in][]{kirk98}.
The loop path has been frequently reported in X-rays \citep[e.g.,][]{sembay93,
takahashi96,kataoka99,zhang99,zhang02,malizia00,ravasio04} and, in one case,
in the infrared \citep{gear86}. In the optical regime, \citet{xilouris06} have
reported a similar pattern, and \citet{wu07a} showed a clockwise loop with
the nightly means of magnitudes and colors. Such hysteresis behavior will
inevitably translate into inter-band time lags.

In our monitoring, there are some mini-flares. The one with the largest
amplitude occurred from 0.245 to 0.292 on JDs~2,454,088 (see Fig.~\ref{F3}).
We checked the color evolution during that flare.
Figure~\ref{F8} displays the result. The numbers associated with the points
signify the evolutionary sequence, with the beginning and end points marked
by the filled circle and square, respectively. There is a hint of a clockwise
loop when we check only the points, but the relatively large errors prevent
us from drawing a convincing conclusion. Therefore, in order to observe an
optical hysteresis loop on the color-magnitude diagram, more accurate
observations are needed, preferentially during a flare with a larger amplitude
than that of the current one. High temporal resolution may also be crucial.

\subsection{Cross Correlation Analyses and Time Lags}
We then performed the inter-band correlation analyses and searched for the
possible inter-band time delay. These are carried out on the intranight
light curves for the four nights with variations, JDs~2,454,088, 2,454,090,
2,454,092 and 2,454,097.

We at first used the z-transformed discrete correlation functions
\citep [ZDCFs;][]{alex97} to search for the $B'$--$V'$, $V'$--$R'$, and
$B'$--$R'$ correlations. The ZDCF differs from the discrete correlation
function (DCF) of \citet{edelson88} in that it bins the data points into
equal population bins and uses Fisher's z-transform to stabilize the highly
skewed distribution of the correlation coefficient. It is much more efficient
than the DCF in uncovering correlations involving the variability timescale,
and deals with under-sampled light curves better than both the DCF and the
interpolated cross-correlation function \citep[ICCF;][]{gaskell87}. For each
correlation, a Gaussian fitting (GF) was made to the central ZDCF points
with ZDCF value greater than $\sim 75\%$ of the peak value. The time
where the Gaussian profile peaks denotes the lag between the correlated light
curves.

The results are listed in Table~4 and displayed in Figures~\ref{F9}, \ref{F10},
\ref{F11}, and \ref{F12} for JDs~2,454,088, 2,454,090, 2,454,092, and
2,454,097, respectively. In each panel, Julian Date and the correlated
passbands ($B'$--$V'$, $V'$--$R'$, or $B'$--$R'$) are given at the upper-left
corner. The peak of the Gaussian profile (the dashed line) is marked with a
vertical dotted line. The corresponding lag is given at the upper-right
corner. A positive lag means that the former variation leads the latter one.

One problem in the GF is that it usually gives an unreliable and
under-estimated error for the lag due at least to the statistically
non-independent ZDCF points. Then, although the lags obtained from the
ZDCF+GF method generally have significance greater than $3\sigma$, they
should be taken with caution. 

Another way to measure the lags and their errors is the ICCF \citep{gaskell87}.
The error was estimated with a model-independent Monte Carlo method, and the
lag was taken as the centroid of the cross-correlation functions that were
obtained with a large number of independent Monte Carlo realizations. This is
the flux-randomization/random-subset-selection (FR/RSS) approach, as described
by \citet{peterson98} and \citet{peterson04}. In the current work, five
thousand independent Monte Carlo realizations were performed on the light
curves on JDs~2,454,088, 2,454,090, 2,454,092, and 2,454,097, and the lags
and errors were obtained and are presented in Table~4. One can see from the
table that although the ZDCF+GF and FR/RSS lags are usually consistent with
each other, the FR/RSS errors are much larger than the corresponding ZDCF+GF
errors. The FR/RSS lags have significance lower than $3\sigma$ except for
the $V'-R'$ and $B'-R'$ lags on JD~2,454,090. On that night, the variability
at the $B'$ and $V'$ bands lead that at the $R'$ band by about 30 minutes.

A few optical lags have been reported in several blazars before.
For example, \citet{romero00} derived a $\sim4$ minute lag of
the $R$ band variations relative to the $V$ band variations for PKS~0537-441
(see their Fig. 5). BL Lac was observed to have its $I$ band variations
delayed by $\sim0.2$ hours with respect to its $B$ band variations
\citep{papa03}. The same authors also reported a lag of $0.9\pm1.0$ hours of
the $I$ band variations relative to the $B$ band variations for S4~0954+658
\citep{papa04}. For S5~0716+714, \citet{qian00} found an upper limit of 6
minutes for the time lag between the $V$ and $I$ band variations. Similarly,
\citet{villata00} reported an upper limit of 10 minutes on the delay of
the $I$ band variations relative to the $B$ band variations. \citet{stalin06}
also presented time lags of 6 and 13 minutes for the $V$ and $R$ variations
on two nights. But they also cautioned about the low significance level of 
their results. Most recently, \citet{poon09} found a possible lag of about 11
minutes between $B$ and $I$ band variations, and \citet{zhang10a} reported
possible lags of a few minutes between low and high optical frequencies.
These previous lags are either upper limits or cautioned by the authors to
have low confidence levels. In the current work, however, the variability at
short wavelengths were observed to lead that at long wavelengths on at least
one night. This positive detection of inter-band time delays presumably has
taken advantage of the simultaneous measurements at three optical wavelengths
of our novel photometric system.

\section{CONCLUSIONS AND DISCUSSIONS}
\subsection{Conclusions}
We monitored S5~0716+714 for 7 nights with our novel photometric system in
2006 December. The photometry was carried out simultaneously at three optical
wavelengths. The object did not show large-amplitude internight variations
during this period, but displayed intranight variations on four nights and
probable variations on one more. Strong BWB chromatism was detected on both
intranight and internight timescales. For the four nights with variations,
the intranight variation amplitude decreases in the wavelength sequence of
$B'$, $R'$, and $V'$. Cross correlation analyses revealed that the variations
at the $B'$ and $V'$ bands led that at the $R'$ band by about 30 minutes
on one night.

\subsection{Discussion on Color Behavior}
S5~0716+714 showed a strong BWB chromatism during this period, either when
the data are considered together or separately for individual nights. Despite
the strong correlation, the points are in fact quite diffuse on the
color-magnitude map (Fig.~\ref{F7}). Data on different nights occupy quite
different regions, and follow distinct color-magnitude correlations. This
dispersion is also evident in \citet{wu07a} (their Fig.~9). As mentioned in
that paper and \citet{dai11}, differences in variation amplitudes and paces
(the lag) can both lead to the observed BWB chromatism. The variation
amplitudes and the possible lags presented in this paper vary from night to
night. So the dispersion on the color-magnitude map may be linked to the
dispersion on the intranight variation amplitudes and/or the possible lags.

For the spectral or color behavior of blazars, different authors have drawn
different conclusions. However, their results on the color-magnitude
correlation of blazars may be biased by many aspects. Here we discuss some
of them.

\begin{enumerate}

\item Non-simultaneous measurements at different wavelengths. Strictly
speaking, the spectral or color index should be calculated with simultaneously
measured fluxes or magnitudes at different wavelengths. However, most
observations can only obtain quasi-simultaneous measurements. For blazars
which do not show significant variations on very short timescales ($\la0.5$
hour), this may not be a problem. However, for those showing very fast
variations, like S5~0716+714, as mentioned in the first
section, this may be a serious problem. The intrinsic brightness of the object
may change measurably when two observations are made successively at separated
wavelengths. These two observations will result in a wrong spectral or color
index. Although an interpolation can ameliorate this problem to some extent,
the irregularity of the variability of blazars makes the interpolation usually
unpractical. Our novel photometric system with the MPIF has the ability to
take measurements simultaneously at three wavelengths. It can thus trace color
changes accurately.

\item Observations taken at multiple telescopes and/or sites. There are likely
systematic differences between data obtained with different telescopes. The
difference can be 0.05 mag or even higher \citep[e.g.,][]{ghise97}. Then the
color index will have an error close to 0.07--0.1 mag, a value comparable
to the scale of the color change. This error will definitely weaken the
color-magnitude correlation.

\item Inclusion of other components of variation. In addition to the
shock-in-jet component, other mechanisms may be involved and dominant for an
episode in blazar variability, such as the geometric effects suggested for
S4~0954+658 \citep{wagner93} and for S5~0716+714 \citep{wu05}. These effects
lead to achromatic variations. In another BL Lac object, OJ~287, a huge
achromatic outburst during 1994-96 was reported by \citet{sillan96}. However,
this outburst was dominated by a component presumably resulting from the
interaction of the primary accretion disk and the secondary black hole in a
binary black hole system \citep{lehto96,valtaoja00,liu02,valtonen08}. When
the {\sl blazar variability} was isolated, it showed a BWB trend \citep{wu06}.

\item Utilization of the averaged fluxes. If, for a blazar, the difference
between variation amplitudes at different wavelengths is very small, and the
spectral change is mainly resulting from time lag, we will not derive a
spectral change when averaging the fluxes within a timescale far longer than
the time lag. Therefore, it is not surprising that \citet{sagar99} and
\citet{stalin06} claimed no spectral change, because they used the averaged
flux or magnitude to derive the spectral or color index. However, when the 
amplitude difference in a blazar dominates the color change, the color change
may also be detected by using averaged fluxes.

\item Utilization of closely separated frequencies. The $B-V$ and $V-R$ color
indices change much less than the $B-R$, $V-I$, and $B-I$ color indices do.
This has been illustrated by \citet{raiteri03} and this work.

\end{enumerate}

There may be other barriers for the detection of the color or spectral changes
and the color-magnitude correlations. These are the five major issues. Some
of them may combine to function in the observations. For example, a campaign
involving several telescopes usually suffer from the first two, as in the case
of \citet{ghise97} and \citet{raiteri03}.

Our photometric system has the advantage of simultaneous measurements at
three optical wavelengths, and suffers none of the above 5 barriers in the
detection of the color change. So the resulting color behavior should have
higher confidence level than those from previous non- or quasi-simultaneous
observations.

\subsection{Discussion on Time Lag}

A few intrinsic or extrinsic models have been proposed to explain the
variability of blazars. The most popular is the shock-in-jet or internal
shock model, in which shocks propagate down the relativistic jet, accelerating
particles and/or compressing magnetic fields, leading to the observed
variability. Depending on the balance between escape, acceleration, and
cooling of the electrons with different energy, either soft (low energy) or
hard (high energy) lags are expected \citep[e.g.,][]{kirk98,sokolov04,
marscher06}. In fact, time lags between variations at much separated
wavelengths are well known. For example, the soft X-ray variations were
observed to lag the hard X-rays by about 1 hour in Mrk~421 \citep{takahashi96}
and by about 1000 s in S5~0716+714 \citep{zhang10b}. In PKS~2155-304, the
X-ray flare appears to lead the EUV and UV fluxes by 1 and 2 days,
respectively \citep{urry97}. Sometimes, the lags may show as the offsets
between the X-ray knots and the radio or optical ones in the relativistic
jets of radio galaxies, with the X-ray knots being closer to the core
\citep[e.g.][]{bai03}.

In optical regime, several authors claimed no detection of inter-band delays
\citep[e.g.][]{hao10,carini11}. Some other claimed the opposite, as mentioned
in \S3.4. Four key parameters probably determine whether the optical time lags
can be detected: (1) wavelength separation, (2) variation amplitude, (3)
temporal resolution, and (4) measurement accuracy. For the first factor, the
larger the wavelength separation, the easier the lag will be detected. Then a
$B-R$ or $B-I$ correlation should be performed rather than a $B-V$ or $V-R$
correlation. For the second, blazars with high duty cycles and large variation
amplitudes should be observed for the search of the lag. For the third, a
temporal resolution of less than 5 minutes is favorable, because most
reported time lags are less than 10 minutes, as mentioned above.
For the fourth, bright sources should be monitored. S5~0716+714 is one of the
brightest blazars in the optical regime. It shows a high variation rate and
has a high duty cycle, as mentioned in \S1. This makes it probably the best
candidate to be monitored for the search of the optical lags.

Our photometric system has the advantage of simultaneous measurements at
multiple wavelengths and increases the temporal resolution significantly in
each single wavelengths. With this system, we detected the time delays on one
night. On the other and, our photometric system has its own deficiencies.
For example, because of the elongated stellar images on the CCD frames,
it is hard to accurately define the center of the aperture, especially for
the $B'$-band images. This results in an extra error in the data reduction.
Another drawback may be that the wavelength separation between the $B'$ and
$R'$ bands is not large enough. Therefore, a photometric system with a light
splitting device may be more suitable in searching for the optical lags. When
achieving a temporal resolution of about or higher than 1 minute, which is
more than an order of magnitude shorter than the timescale of the optical
variability of S5~0716+714 \citep{sasada08,rani10a,chandra11}, the traditional
quasi-simultaneous photometric system may also be able to detect the optical
lags with high confidence level.

\begin{acknowledgments}
We thank the anonymous referee for insightful comments and suggestions that
helped to improve this paper very much. This work has been supported by the
Chinese National Natural Science Foundation grants 10873016, and 11073032,
and by the National Basic Research Program of China (973 Program), No.
2007CB815403.
\end{acknowledgments}

\begin{figure}
\plotone{fig6a.ps}
\plotone{fig6b.ps}
\plotone{fig6c.ps}
\caption{Intranight light curves on JD~2,454,097 in the $B'$, $V'$, and $R'$
bands. The associated small panels give the differential light curves of the
check star (star 6).}
\label{F6}
\end{figure}

\begin{figure}
\epsscale{1.0}
\plottwo{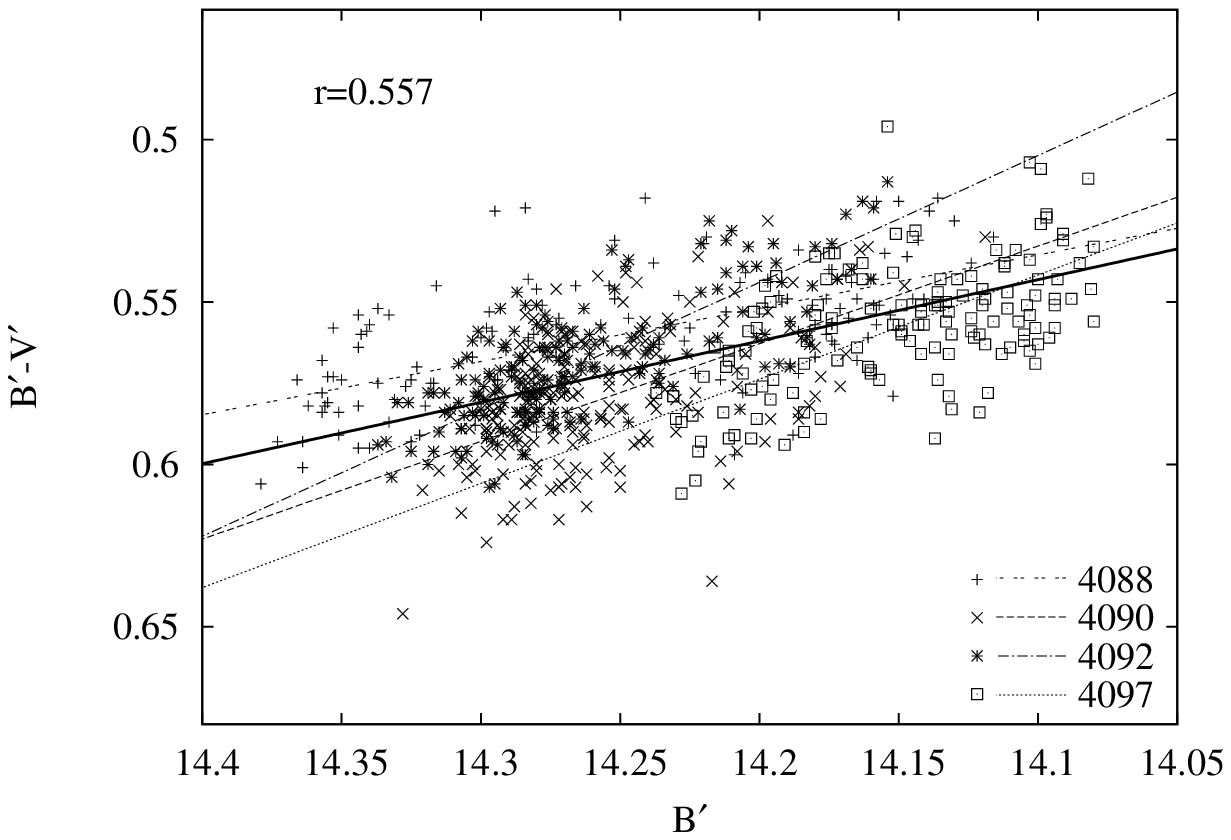}{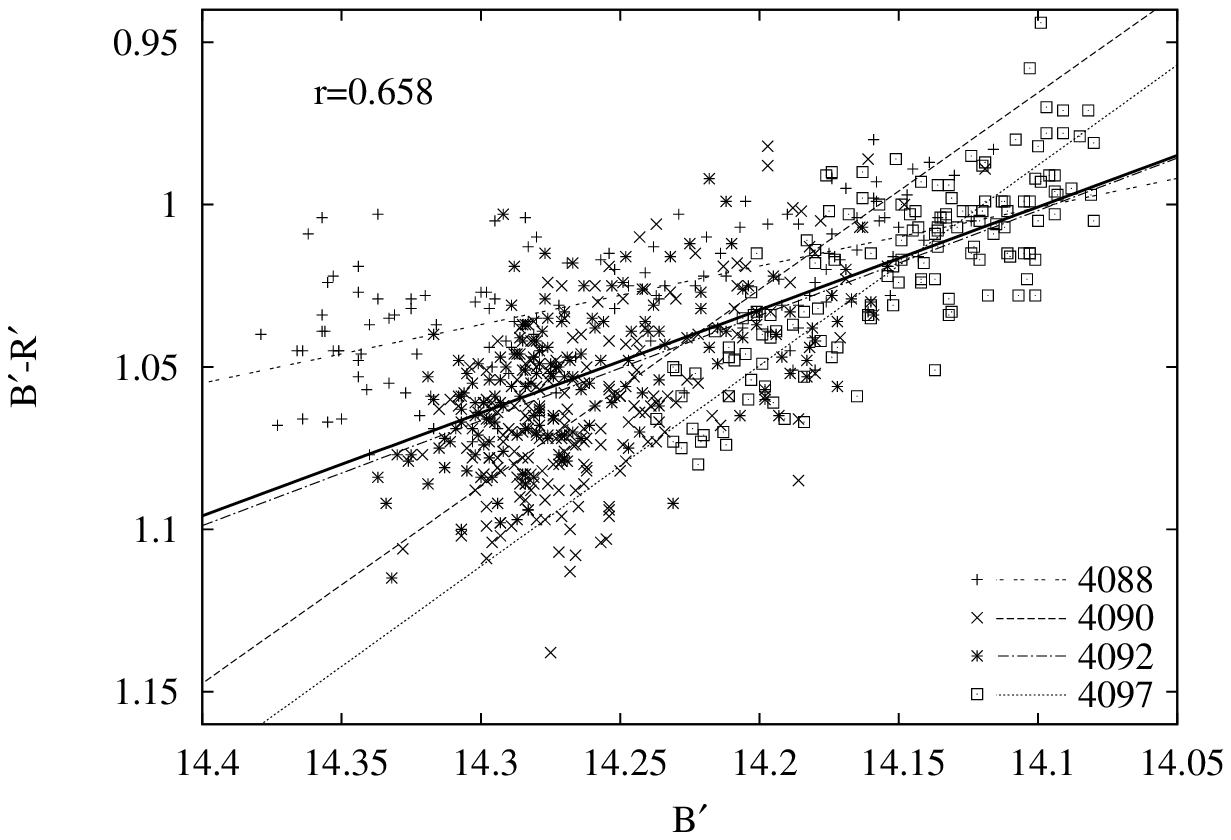}
\caption{Color-magnitude diagrams for JDs ~2,454,088, 2,454,090, 2,454,092,
and 2,454,097. The left panel is for $B'-V'$ color, while the right is for
$B'-R'$ color. Different symbols represent data on different nights. They
are fitted linearly and separately for each individual night and are shown
with different lines. The solid lines are fittings to all points, with the
corresponding correlation coefficients given at the upper-left corners. The
errors are not plotted here for clarity. The mean magnitude error is 0.018
mags and the mean color error is 0.022 mags.}
\label{F7}
\end{figure}

\begin{figure}
\plotone{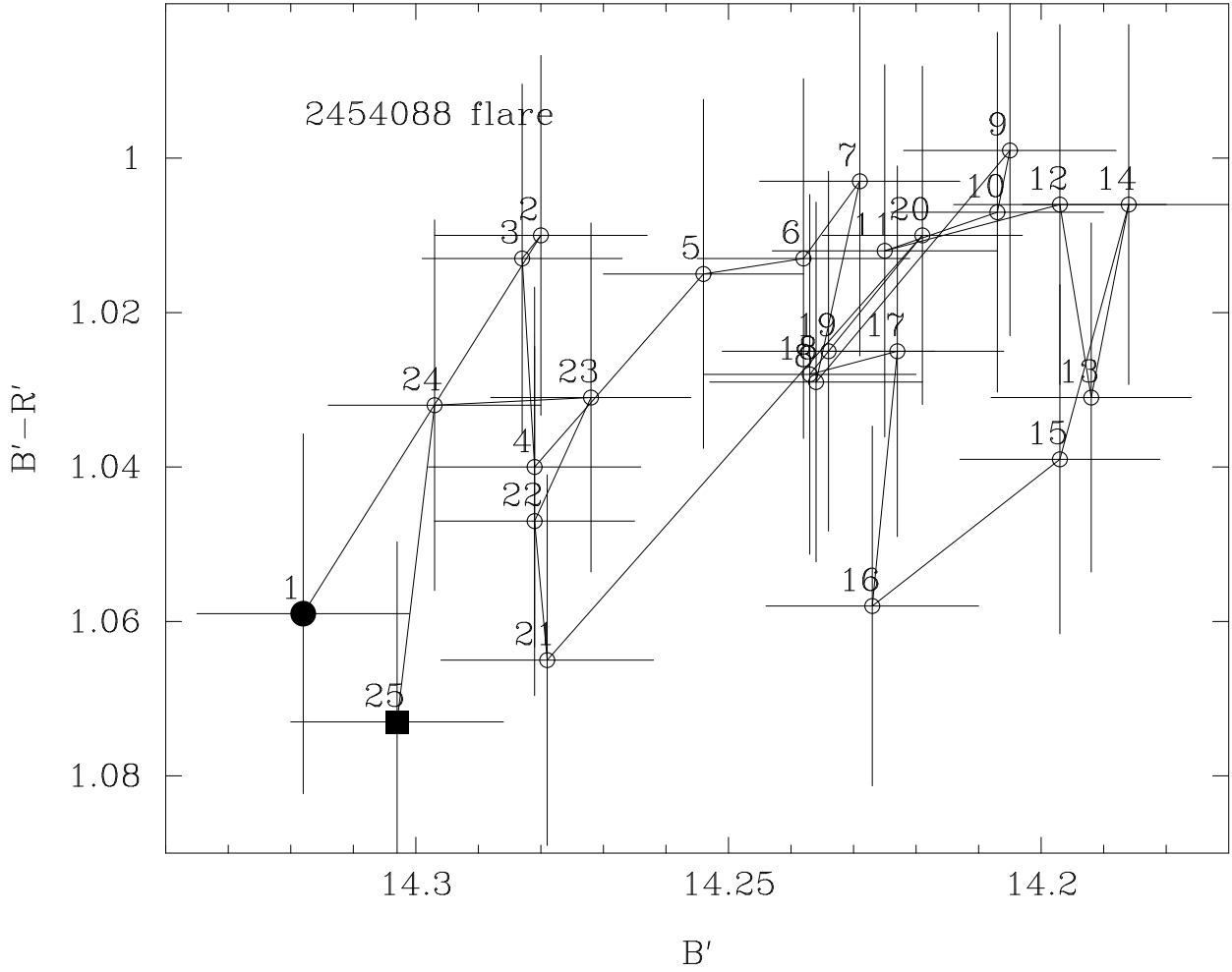}
\caption{Color evolution during flares from 0.245 to 0.292 on JDs 2,454,088.
The filled circle and square mark the beginning and ending points,
respectively. The numbers indicate the points in time sequence.}
\label{F8}
\end{figure}

\begin{figure}
\epsscale{0.5}
\plotone{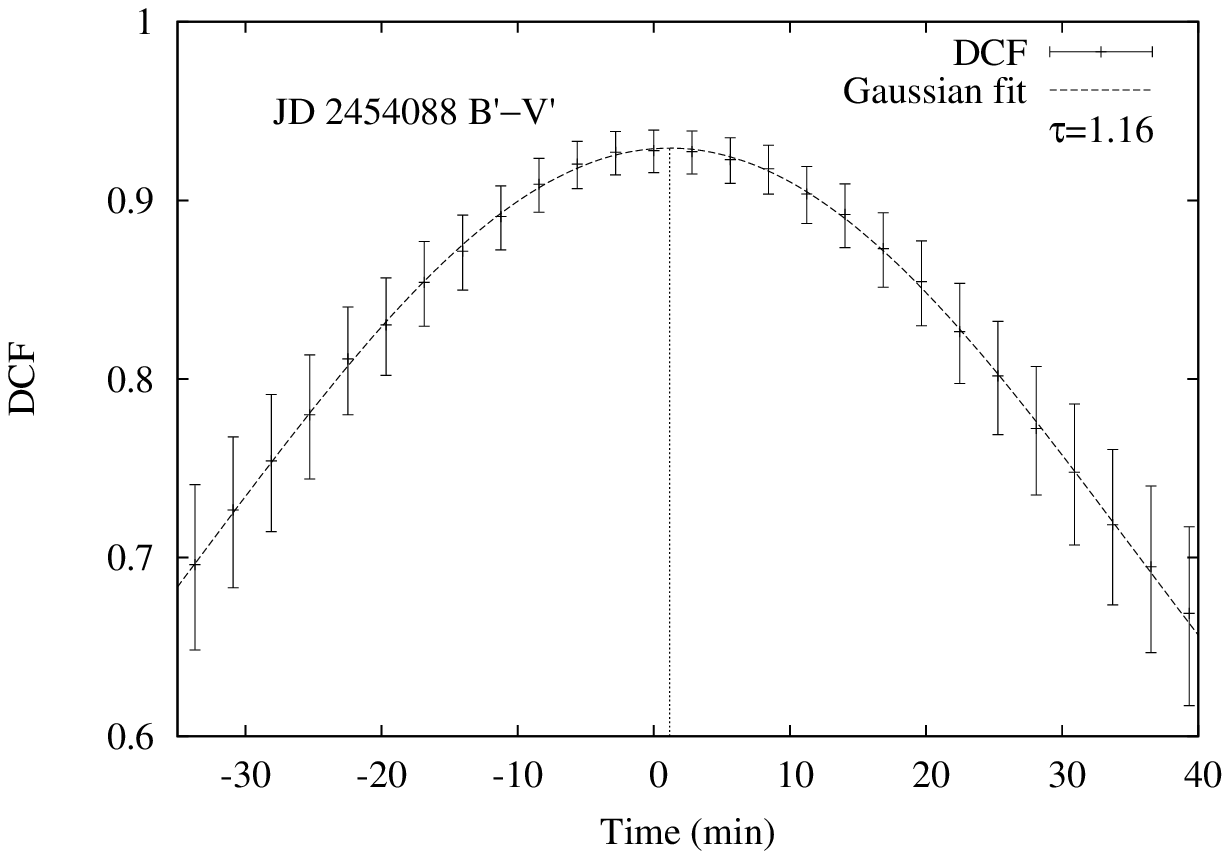}
\plotone{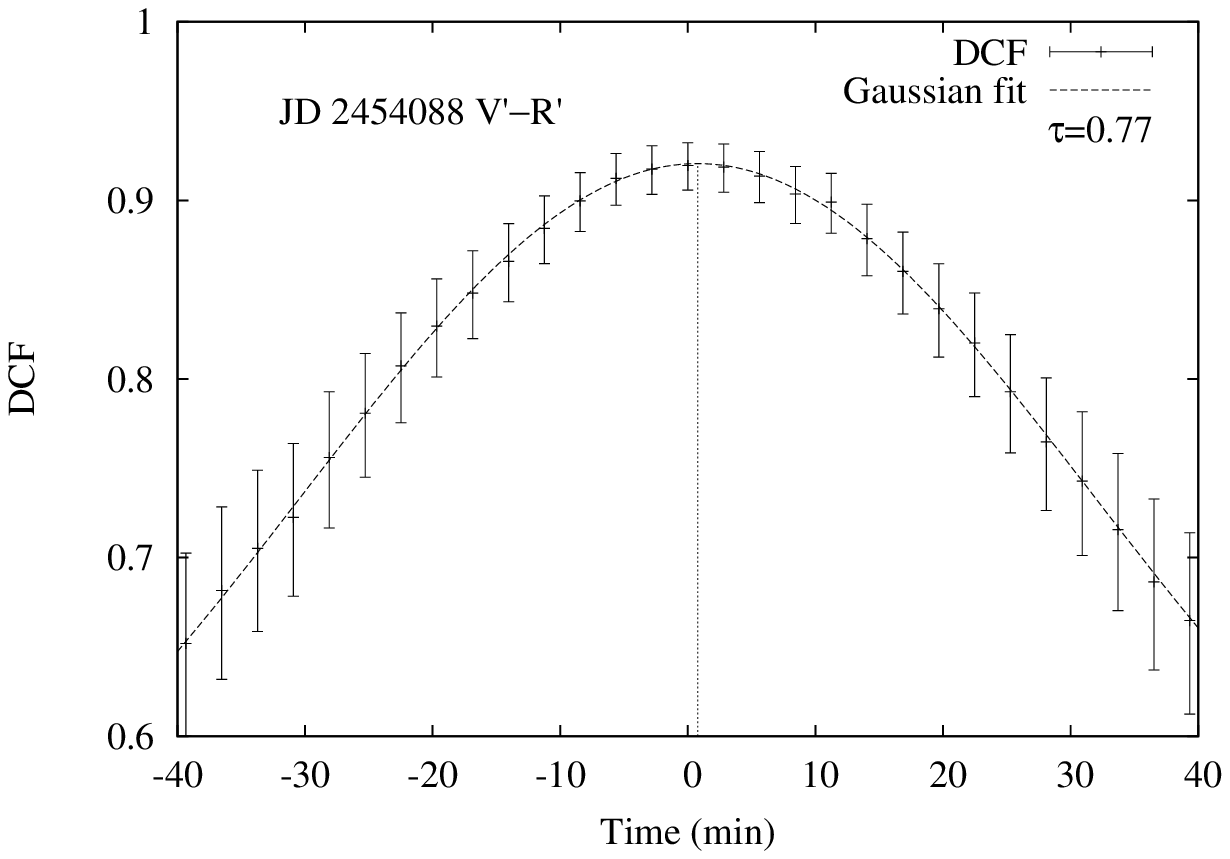}
\plotone{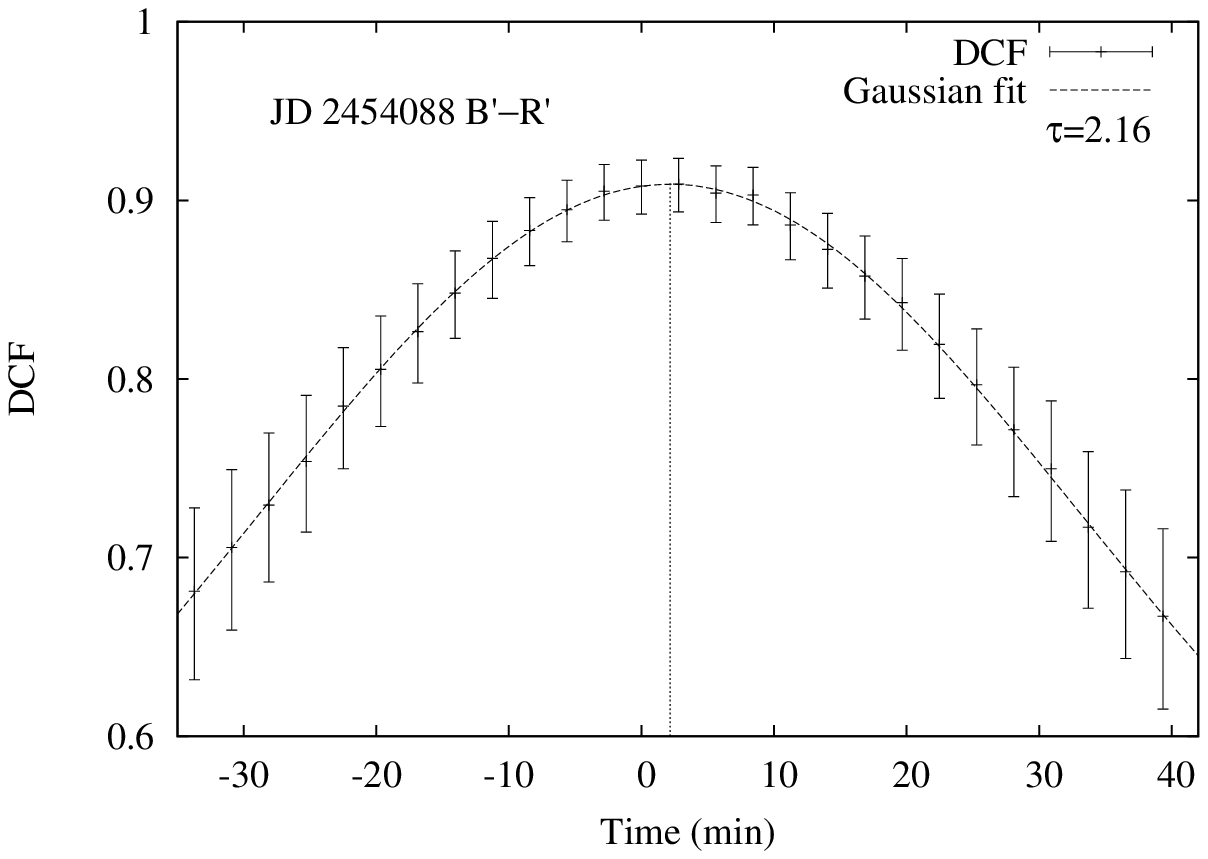}
\caption{ZDCF correlation and fitting results on JD~2,454,088. The dashed
lines are Gaussian fittings to the points, and their peaks are marked with
the vertical dotted lines.}
\label{F9}
\end{figure}

\begin{figure}
\epsscale{0.5}
\plotone{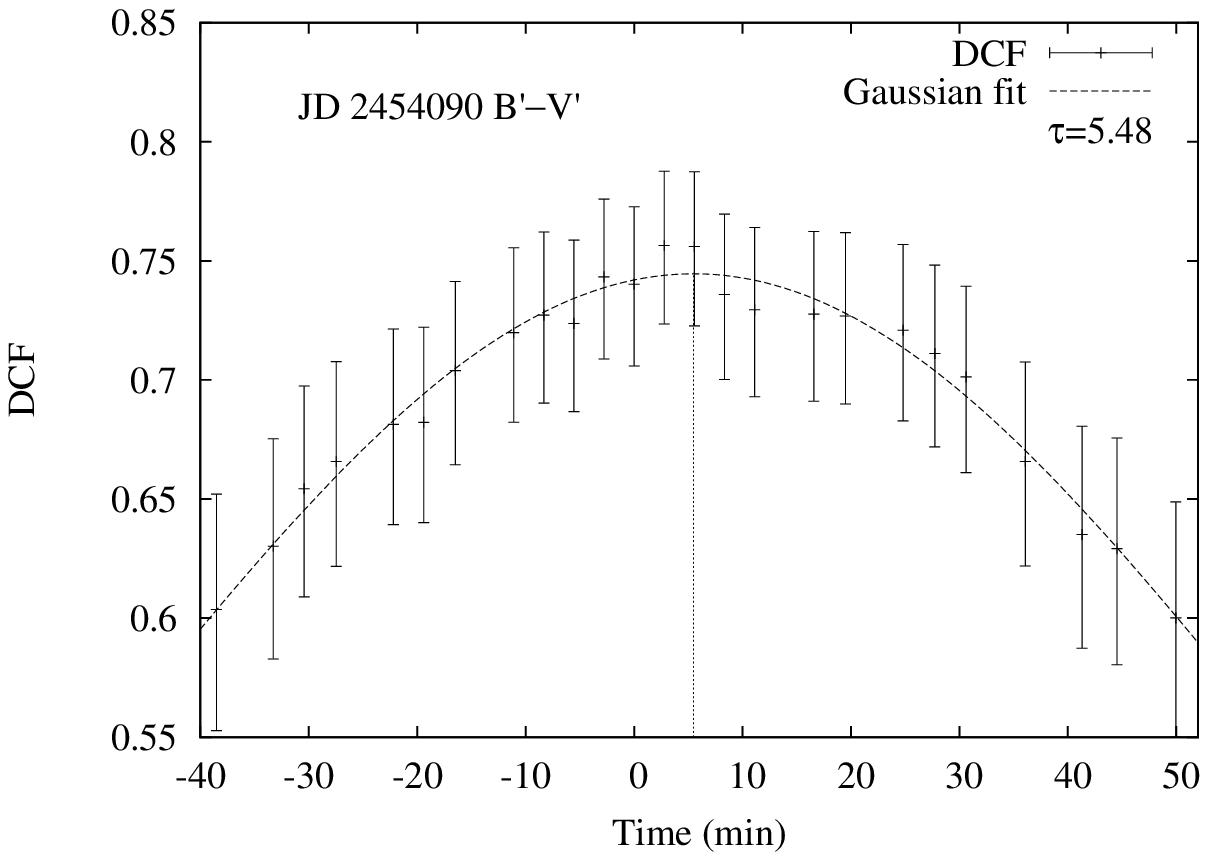}
\plotone{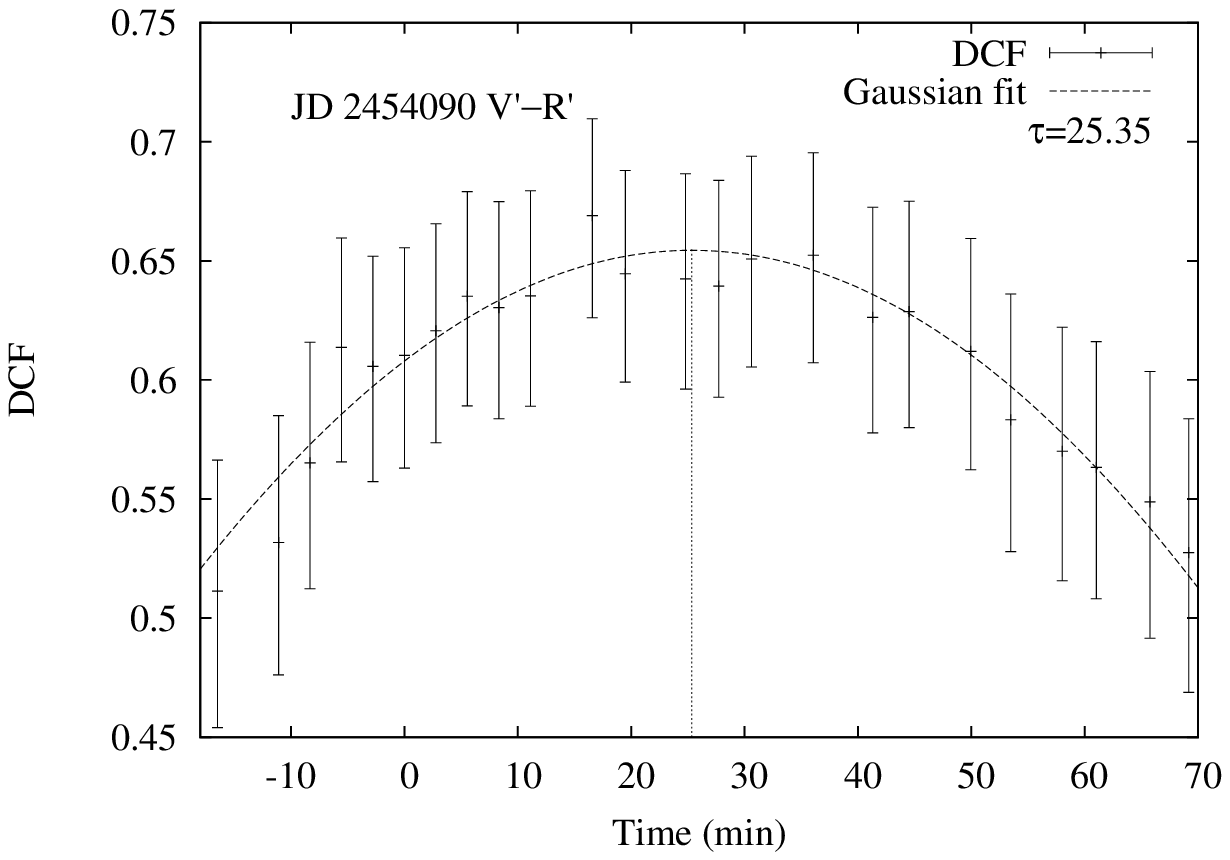}
\plotone{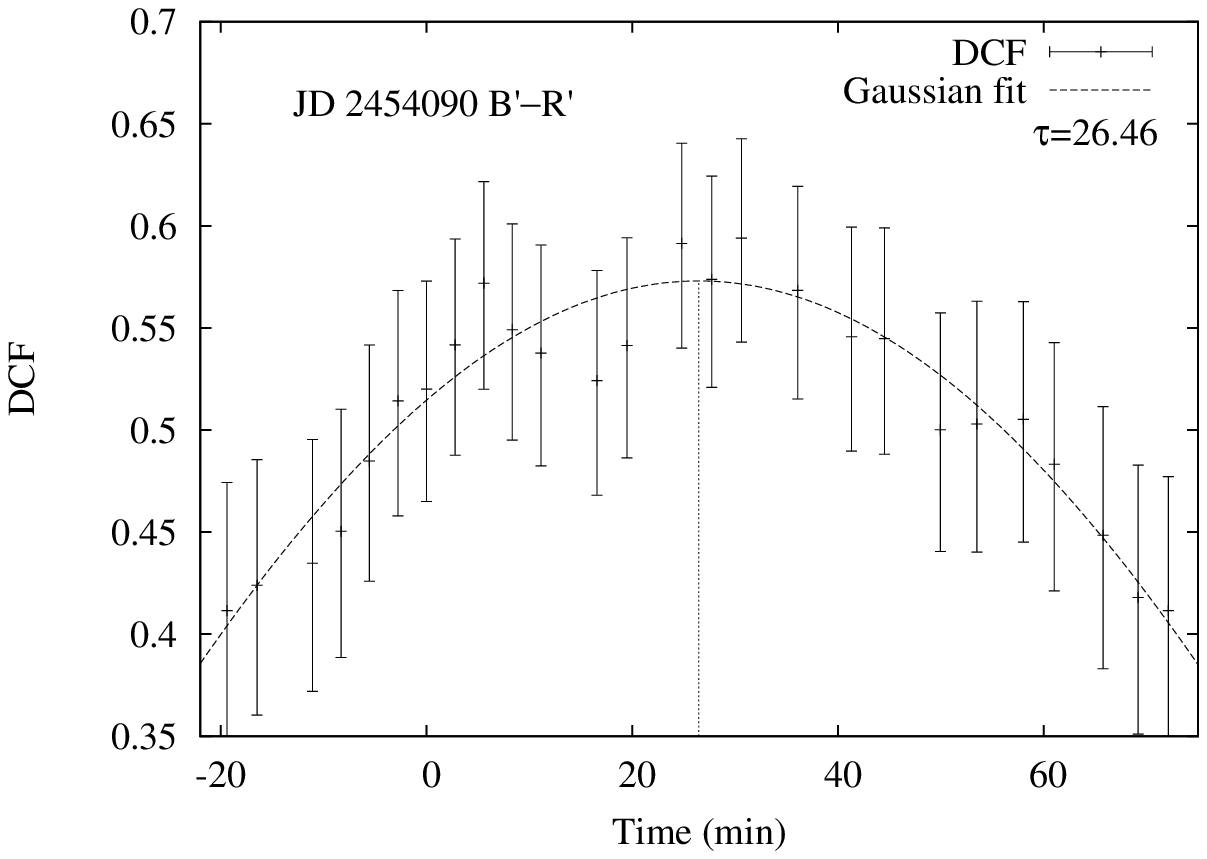}
\caption{ZDCF correlation and fitting results on JD~2,454,090. The dashed
lines are Gaussian fittings to the points, and their peaks are marked with
the vertical dotted lines.}
\label{F10}
\end{figure}

\begin{figure}
\plotone{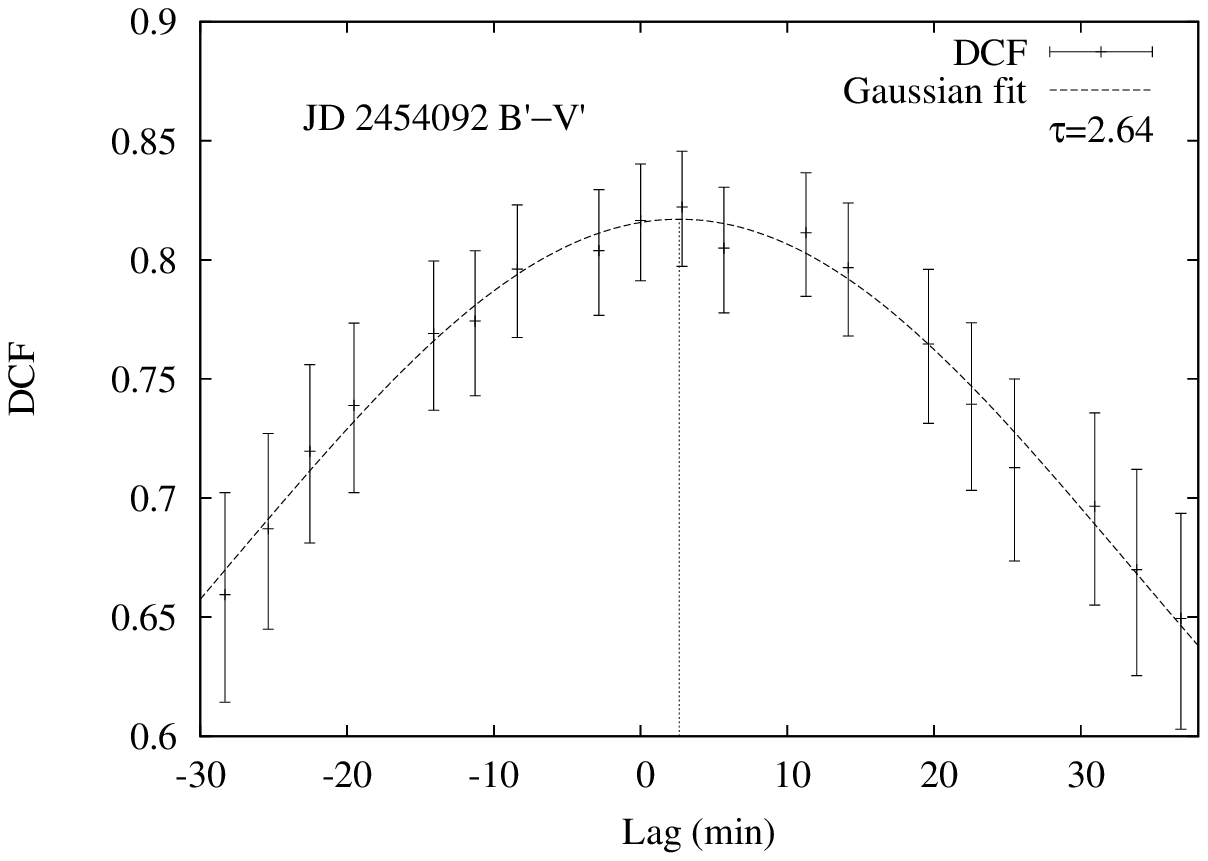}
\plotone{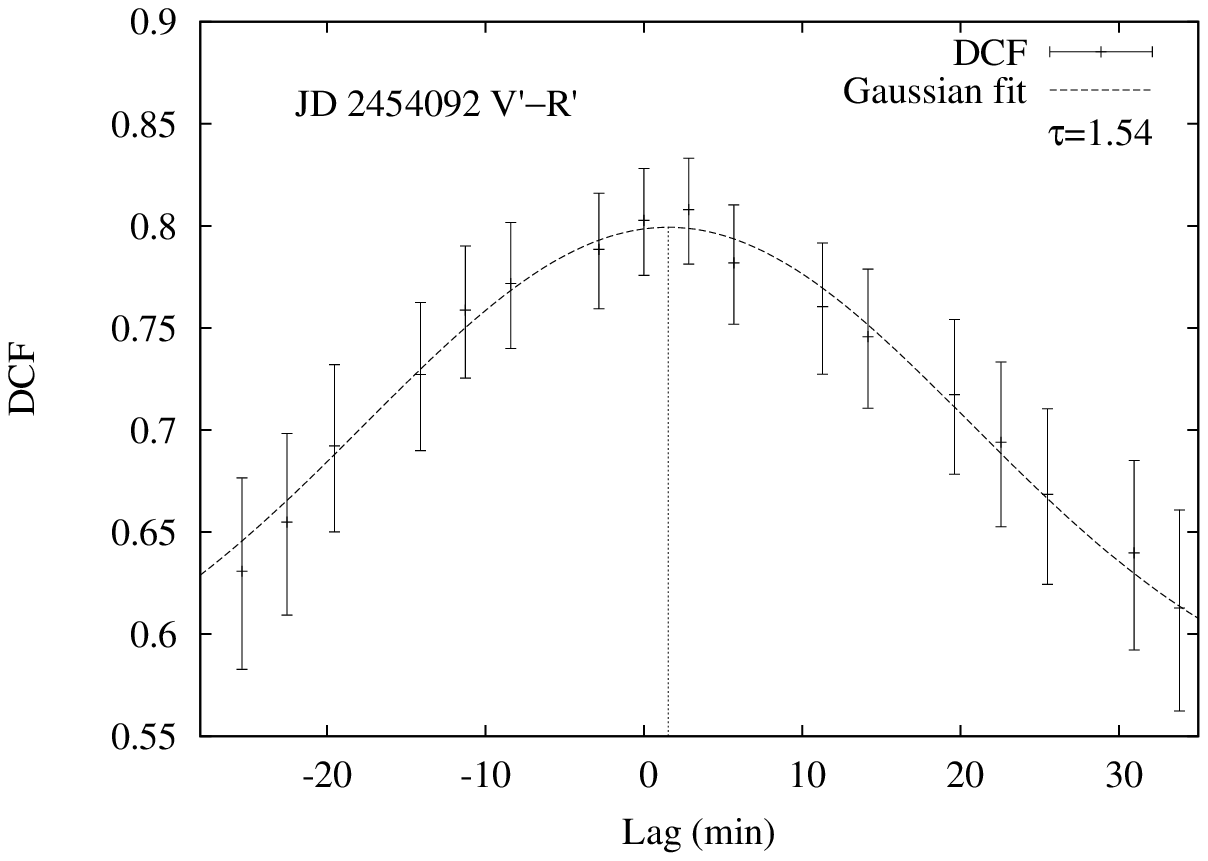}
\plotone{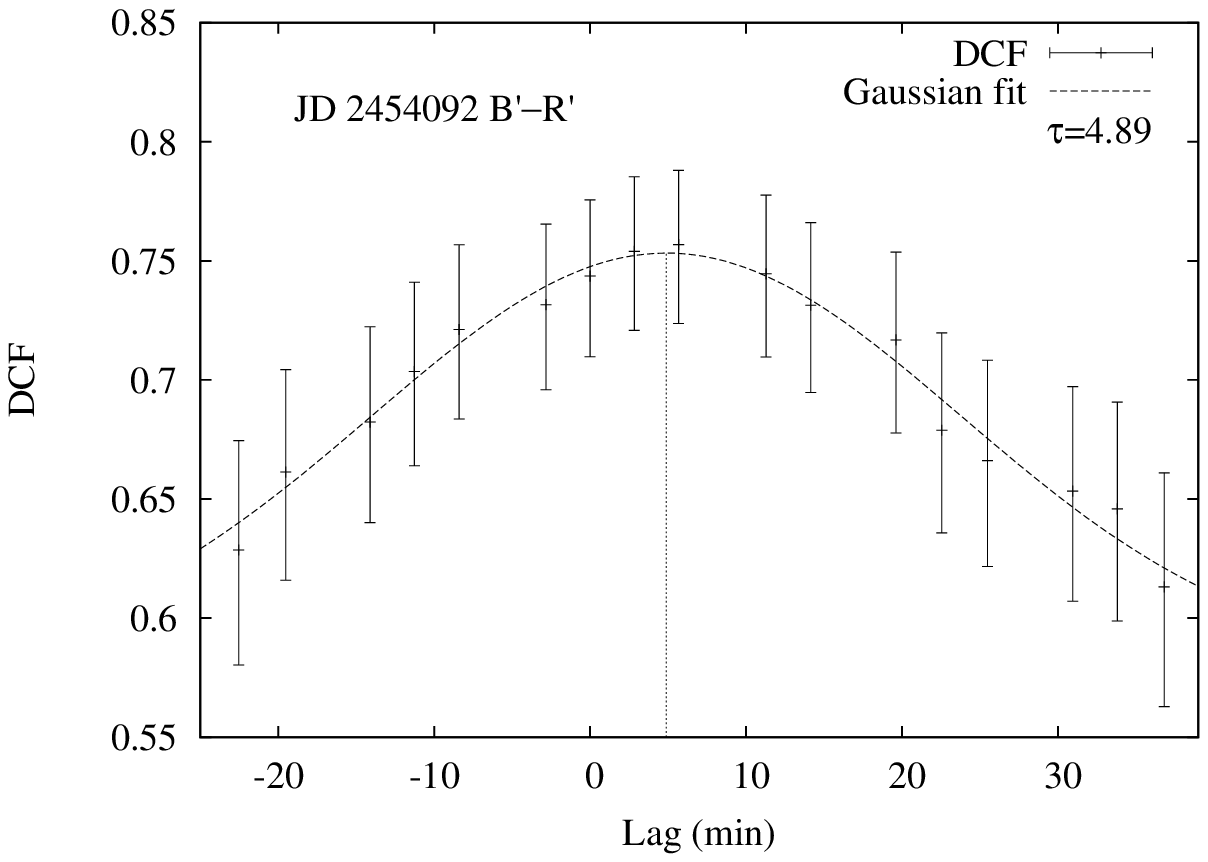}
\caption{ZDCF correlation and fitting results on JD~2,454,092. The dashed
lines are Gaussian fittings to the points, and their peaks are marked with
the vertical dotted lines.}
\label{F11}
\end{figure}

\begin{figure}
\plotone{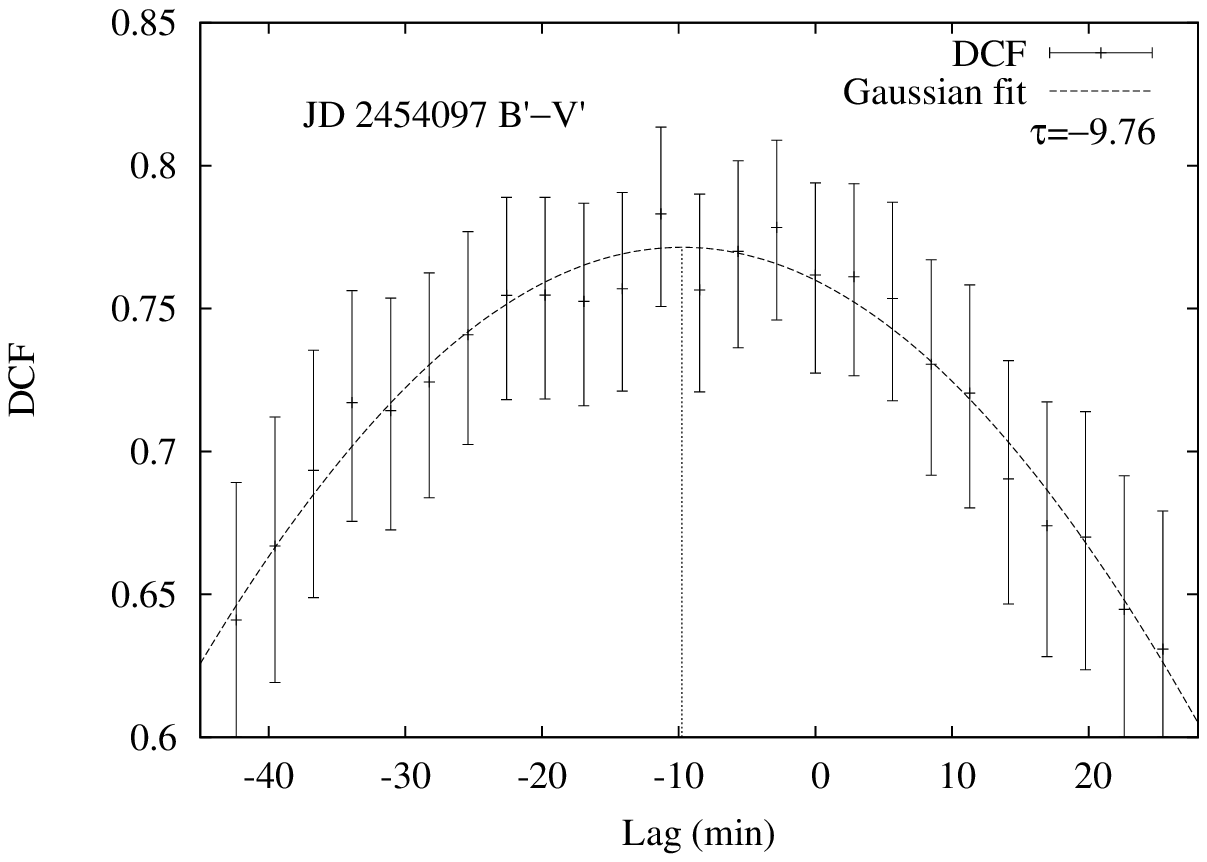}
\plotone{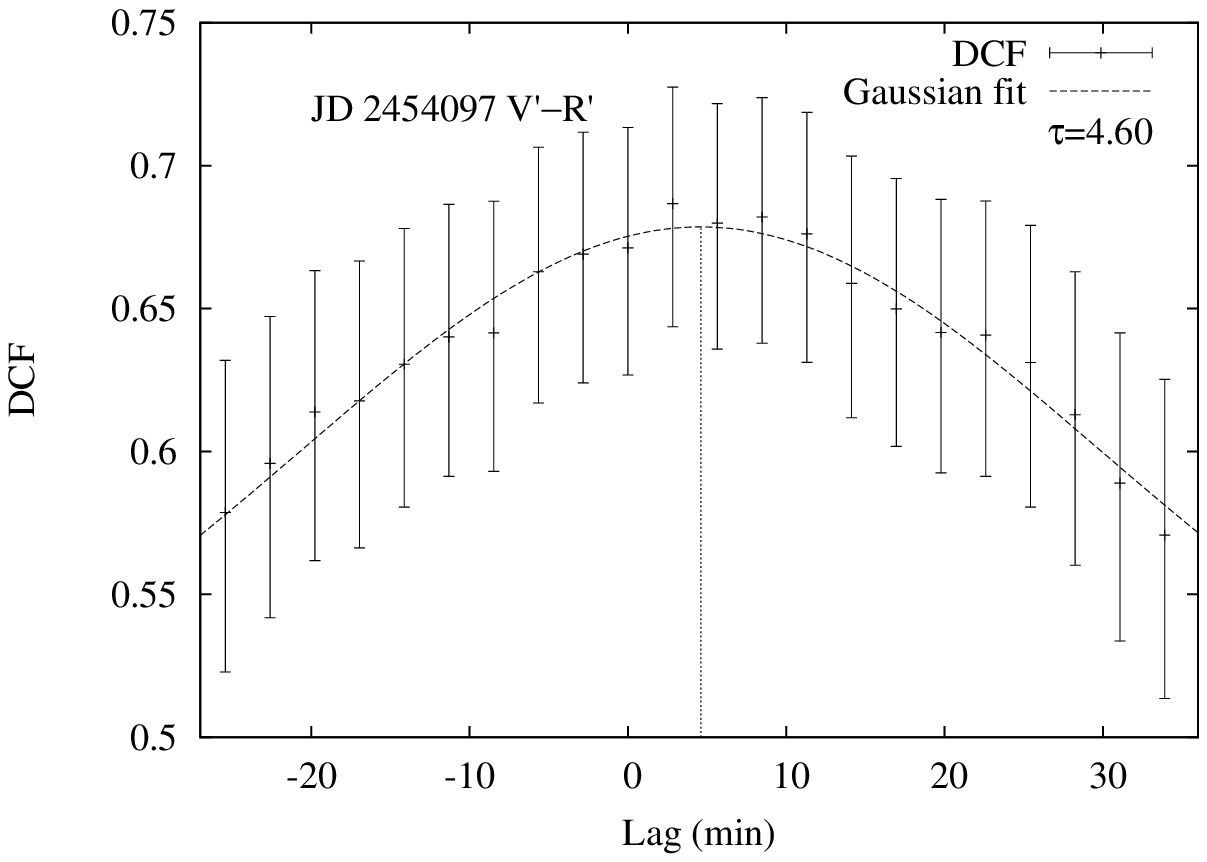}
\plotone{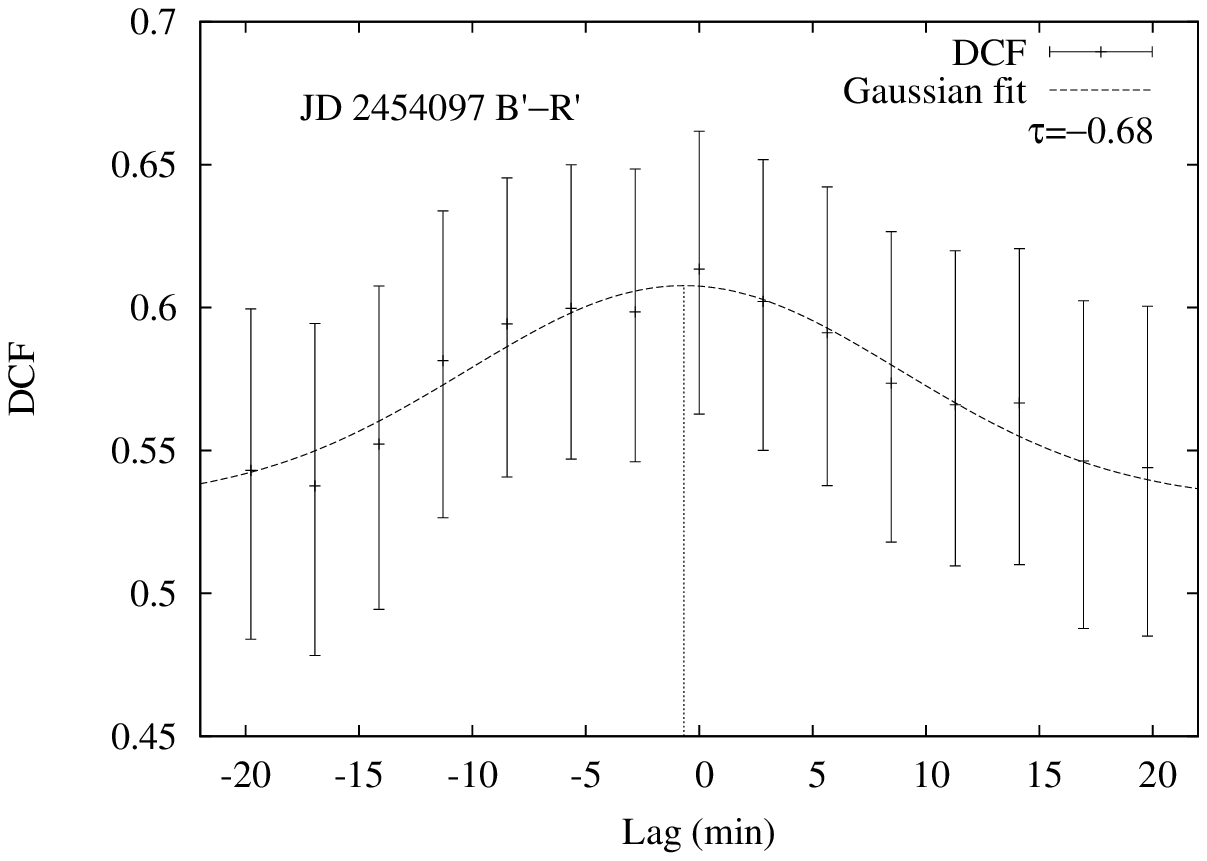}
\caption{ZDCF correlation and fitting results on JD~2,454,097. The dashed
lines are Gaussian fittings to the points, and their peaks are marked with
the vertical dotted lines.}
\label{F12}
\end{figure}

\clearpage

\begin{deluxetable}{ccccccrccrccr}
\rotate
\tablewidth{0pt}
\tablecaption{Observational Log and Results in the $B'$, $V'$ and $R'$ Bands}
\tablehead{\colhead{Date} & \colhead{Time} & \colhead{Julian Date} &
           \colhead{Exp.} & \colhead{$B'$} & \colhead{$B'_{\rm err}$} &
           \colhead{$\delta B'$} & \colhead{$V'$} & \colhead{$V'_{\rm err}$} &
           \colhead{$\delta V'$} & \colhead{$R'$} & \colhead{$R'_{\rm err}$} &
           \colhead{$\delta R'$} \\
           \colhead{(1)} & \colhead{(2)} & \colhead{(3)} & \colhead{(4)} &
           \colhead{(5)} & \colhead{(6)} & \colhead{(7)} & \colhead{(8)} &
           \colhead{(9)} & \colhead{(10)} & \colhead{(11)} & \colhead{(12)} &
           \colhead{(13)}
           }
\startdata
2006 12 18 & 16:03:37.0 & 2454088.16918 & 150 & 14.340 & 0.018 & $-$0.014 & 13.745 & 0.011 &    0.008 & 13.263 & 0.017 &    0.014 \\
2006 12 18 & 16:06:25.0 & 2454088.17112 & 150 & 14.298 & 0.017 & $-$0.000 & 13.740 & 0.011 & $-$0.021 & 13.271 & 0.017 & $-$0.015 \\
2006 12 18 & 16:09:14.0 & 2454088.17308 & 150 & 14.325 & 0.018 & $-$0.025 & 13.751 & 0.012 & $-$0.011 & 13.293 & 0.017 & $-$0.005 \\
2006 12 18 & 16:12:02.0 & 2454088.17502 & 150 & 14.320 & 0.017 & $-$0.004 & 13.738 & 0.011 & $-$0.009 & 13.292 & 0.017 & $-$0.010 \\
2006 12 18 & 16:14:51.0 & 2454088.17698 & 150 & 14.316 & 0.017 &    0.011 & 13.771 & 0.011 & $-$0.011 & 13.279 & 0.017 & $-$0.001 \\
\enddata
\tablecomments{Table 1 is published in its entirety in the electronic edition
of the {\sl Astronomical Journal}. A portion is shown here for guidance
regarding its form and content.}
\end{deluxetable}

\begin{deluxetable}{cccrrrcl}
\tablewidth{0pt}
\tablecaption{Variation Test Result}
\tablehead{\colhead{JD} & \colhead{$\nu_1$} & \colhead{$\nu_2$} &
           \colhead{$F_{\rm B'}$} & \colhead{$F_{\rm V'}$} &
           \colhead{$F_{\rm R'}$} & \colhead{$F_{0.01}$} & \colhead{Var?}
          }
\startdata
2,454,088 & 26 & 108 & 23.705 & 29.534 & 25.166 & 1.930 & Y \\
2,454,090 & 36 & 148 &  4.233 & 16.408 &  3.600 & 1.763 & Y \\
2,454,091 & 37 & 152 &  1.599 &  5.270 &  2.817 & 1.750 & Y? \\
2,454,092 & 36 & 148 &  8.342 & 13.132 &  8.296 & 1.763 & Y \\
2,454,093 & 38 & 156 &  0.687 &  2.584 &  1.317 & 1.740 & N \\
2,454,094 & 13 &  56 &  0.684 &  0.523 &  1.521 & 2.465 & N \\
2,454,097 & 38 & 156 &  4.359 & 11.649 &  3.068 & 1.740 & Y \\
\enddata
\end{deluxetable}

\begin{deluxetable}{cccc}
\tablewidth{0pt}
\tablecaption{Intranight Variation Amplitude}
\tablehead{ & \colhead{$A_{\rm B'}$} & \colhead{$A_{\rm V'}$} &
            \colhead{$A_{\rm R'}$} \\
           \colhead{JD} & (mag) & (mag) & (mag) 
          }
\startdata
2,454,088 & 0.262 & 0.222 & 0.233 \\
2,454,090 & 0.208 & 0.148 & 0.152 \\
2,454,092 & 0.182 & 0.132 & 0.172 \\
2,454,097 & 0.206 & 0.134 & 0.110 \\
\enddata
\end{deluxetable}

\begin{deluxetable}{crrcrrcrr}
\tablewidth{0pt}
\tablecaption{Time Lags Between Different Wavelengths\tablenotemark{a}}
\tablehead{ & \multicolumn{2}{c}{$B'$--$V'$} & & \multicolumn{2}{c}{$V'$--$R'$}
      & & \multicolumn{2}{c}{$B'$--$R'$} \\
      \cline{2-3} \cline{5-6} \cline{8-9}
      JD & \colhead{ZDCF+GF} & \colhead{FR/RSS} & \colhead{} & \colhead{ZDCF+GF} &
      \colhead{FR/RSS} & \colhead{} & \colhead{ZDCF+GF} & \colhead{FR/RSS}
      }
\startdata
 2,454,088 & 1.16$\pm$0.08 & 1.26$\pm$1.92 & & 0.77$\pm$0.08 & 0.09$\pm$2.04 & & 2.16$\pm$0.08 & 2.40$\pm$2.16 \\
 2,454,090 & 5.48$\pm$0.51 & 4.98$\pm$5.82 & & 25.35$\pm$0.81 & 28.14$\pm$7.02 & & 26.46$\pm$0.92 & 30.12$\pm$11.04 \\
 2,454,092 & 2.64$\pm$0.37 & 2.52$\pm$3.48 & & 1.54$\pm$0.40 & 3.30$\pm$2.94 & & 4.89$\pm$0.46 & 6.78$\pm$4.26 \\
 2,454,097 & $-$9.76$\pm$0.47 & $-$10.2$\pm$4.38 & & 4.60$\pm$0.40 & 3.54$\pm$5.76 & & $-$0.68$\pm$0.56 & $-$4.86$\pm$9.66 \\
\enddata
\tablenotetext{a}{The lags are in unit of minute.}
\end{deluxetable}

\end{document}